%% file: biblatex-plgd-virtualworld.tex
\documentclass[a4paper,12pt]{article}
\usepackage{geometry}
\pdfoutput=1

\usepackage[T1]{fontenc} %
\usepackage[swedish,american]{babel}

\usepackage[autostyle]{csquotes}

\usepackage[backend=bibtex,style=authoryear-ibid,backref=false,uniquename=false]{biblatex} %
\makeatletter

\newrobustcmd*{\parentexttrack}[1]{%
  \begingroup
  \blx@blxinit
  \blx@setsfcodes
  \blx@bibopenparen#1\blx@bibcloseparen
  \endgroup}

\AtEveryCite{%
  }

\makeatother

\let\cite\parencite
\let\citeN\textcite
\usepackage{hyperref}
\usepackage{adjustbox} %
\usepackage{relsize} %
\usepackage{url}
\def\UrlFont{\small\tt} %
\usepackage{amssymb} %
\usepackage{amsthm}
\theoremstyle{definition}
\newtheorem{definition}{Definition}[section]
\usepackage{paralist}
\usepackage{rotating} %
\usepackage{pifont}
\usepackage{wrapfig}
\usepackage{mdframed}
\usepackage{footnote}

\usepackage{afterpage}

\usepackage[shadow,textsize=small,disable]{todonotes}
\newcommand\todoin[2][]{\todo[inline, caption={2do}, #1]{
	\begin{minipage}{\textwidth-4pt}#2\end{minipage}}}

\usepackage{xspace}
\newcommand{\ie}{\textit{i.e.,}\xspace}
\newcommand{\eg}{\textit{e.g.,}\xspace}
\newcommand{\sic}{\textit{sic}}

\newcommand{\REQUIRES}{\medskip\noindent\textsc{REQUIRES:}\xspace}

\newcommand{\PLUS}{\texttt{+}\xspace}
\newcommand{\ZERO}{\textsc{zero}\xspace}
\newcommand{\ONE}{\textsc{one}\xspace}
\newcommand{\MANY}{\textsc{many}\xspace}

\newcommand{\Y}{\ding{51}\xspace} %
\newcommand{\N}{\ding{55}\xspace} %
\newcommand{\R}{-\xspace}
\newcommand{\EQ}{$\vDash$\xspace}
\usepackage{mathabx}
\newcommand{\AND}{$\land$}
\newcommand{\XOR}{$\veebar$} %

\newcommand{\ISA}{\textsc{is-a}\xspace}
\newcommand{\HASA}{\textsc{has-a}\xspace}

\newcommand{\VE}{\texttt{VE}\xspace}
\newcommand{\ST}{\texttt{ST}\xspace}
\newcommand{\VT}{\texttt{VT}\xspace}
\newcommand{\OT}{\texttt{1T}\xspace}
\newcommand{\Rt}{\texttt{Rt}\xspace}
\newcommand{\VS}{\texttt{VS}\xspace}
\newcommand{\OS}{\texttt{1S}\xspace}
\newcommand{\Sh}{\texttt{1Sh}\xspace}
\newcommand{\OH}{\texttt{1H}\xspace}
\newcommand{\TwoH}{\texttt{$2\PLUS$H}\xspace}
\newcommand{\HA}{\texttt{HA}\xspace}
\newcommand{\SA}{\texttt{SA}\xspace}
\newcommand{\Ix}{\texttt{Ix}\xspace}
\newcommand{\nZ}{\texttt{nZ}\xspace}
\newcommand{\wP}{\texttt{wP}\xspace}
\newcommand{\dP}{\texttt{dP}\xspace}
\renewcommand{\P}{\texttt{P}\xspace} %
\newcommand{\Av}{\texttt{Av}\xspace}

\xspaceaddexceptions{\AND\XOR[]\}}

\newcommand{\sectioncaps}[1]{ \section{#1} } %

\begin{document}

\title{Virtual World, Defined from a Technological Perspective, and Applied to Video Games, \mbox{Mixed Reality} and the Metaverse [v-0.16]}
\author{Kim J.L. Nevelsteen\\
Immersive Networking, 
Dept. of Computer and Systems Sciences, \\
Stockholm University
}
\date{}
\maketitle

\input{incl/0-abstract}

\input{incl/1-introduction}

\input{incl/2-relatedwork}

\input{incl/4-approach}

\input{incl/results}

\input{incl/5-results}

\input{incl/6-verification}

\input{incl/7-conclusion}

\section*{\textsc{Acknowledgements}}
Sincerest gratitude to Theo Kanter and Rahim Rahmani for their guidance.
Thank you to Magnus Johansson, Henrik Warpefelt and Jim Wilenius, for their intellectual input.
This research was made possible by a grant from the Swedish Governmental Agency for Innovation Systems to the Mobile Life Vinn Excellence Center.

\renewcommand*{\bibfont}{\small}
\renewcommand*{\UrlFont}{\ttfamily\smaller\relax}

\printbibliography[title=\textsc{References},nottype=technology]

\input{incl/Appendix}

\defbibfilter{technologies}{%
  type=technology
}
\printbibliography[title=\textsc{Technologies},filter=technologies]

\end{document}

%% file: incl/0-abstract.tex
\newcommand{\DOMAIN}{}
\newcommand{\RQ}{}
\newcommand{\PROBLEM}{}
\newcommand{\RELATEDWORK}{}
\newcommand{\APPROACH}{}
\newcommand{\VERIFICATION}{}
\newcommand{\CONCLUSION}{}
\newcommand{\IMPLICATIONS}{}

\begin{abstract}
There is no generally accepted definition for a virtual world, with many complimentary terms and acronyms having emerged implying a virtual world. 
Advances in networking techniques such as, host migration of instances, mobile ad-hoc networking, and distributed computing, bring in to question whether those architectures can actually support a virtual world.
Without a concrete definition, controversy ensues and it is problematic to design an architecture for a virtual world. 
Several researchers provided a definition but aspects of each definition are still problematic and 
simply can not be applied to contemporary technologies. 
The approach of this article is to sample technologies using grounded theory, and obtain a definition for a `virtual world' that is directly applicable to technology. 
The obtained definition is compared with related work and used to classify advanced technologies, such as: a pseudo-persistent video game, a MANet, virtual and mixed reality, and the Metaverse.
The results of this article include: a break down of which properties set apart the various technologies; a definition that is validated by comparing it with other definitions; an ontology showing the relation of the different complimentary terms and acronyms; and, the usage of pseudo-persistence to categories those technologies which only mimic persistence.
\end{abstract}

%% file: incl/1-introduction.tex
\sectioncaps{Introduction}

\PROBLEM There is no generally accepted definition for a virtual world and \citeN{schroeder2008-defining} highlights the need for a definition. 
There is controversy on the meaning of the term `virtual world' in various communities and research~\cite{combs2004-def,brennan2009-pesky}, with many complimentary terms and acronyms emerging implying a virtual world, but which are qualified %
so as to specialize the concept \eg MMORPG. 
Advances in networking such as, host migration of instances, mobile ad-hoc networking, and distributed computing, bring into question whether architectures can actually support a virtual world. With concepts, such as virtual reality~\cite{coquillart2011-vr} and the Metaverse~\cite{dionisio2013-metaverse}, should these contemporary architectures be classified as virtual worlds or something else?
When a virtual world is not well defined and without a concrete definition, designing an architecture specifically for a virtual world is problematic \eg tactics often used in existing publications are: refraining from providing a reference to the virtual environment used, naming properties that constitute the virtual environment relative to their publication, or only referring to examples of existing implementations. %

\RELATEDWORK \citeN{bell2008-virtualworld} has published a think piece to spur up discourse towards a definition. \citeauthor{bartle2003} initially described properties of a virtual world in his book, Designing Virtual Worlds~\cite*{bartle2003}, and later updated the definition, clarifying each individual property. Several other researchers~\cite{bell2008-expanded,girvan2013-virtualworld,spence2008-demographics} have also attempted to provide a definition (\eg through literature review), but aspects of each definition are still problematic. Current definitions simply can not be applied to contemporary technologies, allowing for the classification of technologies such as: the online productivity tool, Google Docs~\cite{google2007-docs}; the video game, League of Legends (LoL)~\cite{riot2009-lol}; the many shards of the virtual world, World of Warcraft (WoW)~\cite{blizzard2004-wow}; or, the Internet as a whole. It doesn't buy us anything to refer to everything as a virtual world, even though it might be possible.

\APPROACH If a definition for a virtual world can be determined, it can be used as a base which can then be qualified to more specific usages \eg qualifying the definition for a networked virtual environment (net-VE)~\cite{singhal1999-netve} with the `massive multiplayer' property to form a type of MMVE. 
Rather than obtain a definition for a virtual world through a literature review, the approach in this article is to use grounded theory~\cite{johannesson2014-designscience}. Sample technologies are added to the study and analyzed for properties related to a virtual world; properties that are determinants of a virtual world together form the new definition. The analysis continues until the definition is no longer refined, but sampling simply confirms the theory. Twenty-six technology samples were analyzed including commonly used software, communication technologies, video games, web technologies, and early and contemporary virtual worlds.
The properties found in \cite{bell2008-virtualworld} and \cite{girvan2013-virtualworld}, provided a starting point for the study.

\VERIFICATION The resulting definition, obtained through the grounded theory, is verified against prominent existing definitions. The effectiveness of the definition is demonstrated by creating an ontology of virtual worlds \ie clarifying acronyms related to virtual worlds \eg VE, MUVE, MMVE, VI3DW, MMOW, MMORPG or IVW.
And, a form of `discriminant sampling'~\cite{creswell2013-qualitative} is used \ie selecting advanced contemporary technologies that were not in the study to see if the theory holds true for these additional samples; classifying advanced technologies, such as: a pseudo-persistent video game, a MANet, virtual reality, mixed reality, and the Metaverse.

\CONCLUSION Results of this article include a detailed definition for a `virtual world', with detailed definitions of all underlying terms used. The definition is applied directly to technology for classification, showing which properties set apart the different technologies. Remaining properties which do not determine a virtual world, are listed so that they can be used to distinguish between the various virtual worlds. 
The resulting definition is tied to the accepted definition of a net-VE, by \citeN{singhal1999-netve}.
Distributed peer-to-peer, cloud-based and MANet networking are taken into account so that they don't invalidate the definition. And, the concept of pseudo-persistence~\cite{nevelsteen2016-persistence,soderlund2009-proximity} is used to categorize technologies which only mimic persistence.

%% file: incl/2-relatedwork.tex
\sectioncaps{Related Work}

Many descriptions and definitions for a virtual world can be found in literature, but there is no consensus. \citeN{dionisio2013-metaverse} state that ``virtual worlds are persistent online computer-generated environments where multiple users in remote physical locations can interact in real time for the purposes of work or play'', but the description assumes a virtual world accessed by a single player on a local computer, rather than from a remote location, fails to be a virtual world, even if the implementation is unchanged. 

\citeauthor{bartle2010-history} \cite*{bartle2010-history} updated his definition with respect to his previous publication \cite{bartle2003}. 
Unfortunately, the updated definition only implies shared time and space, and fails to capture the `worldliness' of a virtual world. Each property found in his definitions is added to the study below and his concepts of an avatar and persistence are adopted.

\citeN{bell2008-expanded} obtain a definition through literature review, stating a virtual world to be ``a synchronous, persistent network of people, represented as avatars, [and] facilitated by networked computers''. 
Aspects of this gathered definition are not sound: synchronous communication is mentioned, but the definition doesn't take into account if participants share time or space; similar to \citeN{dionisio2013-metaverse}, a single player on a local environment fails to be in a virtual world; and the concept of agency is discussed with regards to avatars, but is not captured in the definition. \citeauthor{bell2008-expanded} state that social networks like Facebook are not virtual worlds, because ``a social networking site has persistence but no sense of synchronous environment (and therefore no sense of space)''. In that statement, a synchronous environment is unclearly attributed to a sense of space. The underlying definition used, by Raph Koster~\cite[\#33]{combs2004-def}, stating that a virtual world is ``a spatially based depiction of a persistent virtual environment, which can be experienced by numerous participants at once, who are represented within the space by avatars'', is of interest; the properties of spatial depiction, persistence and avatars are considered in the study below.

Similar to \citeauthor{bell2008-expanded}, \citeN{girvan2013-virtualworld}\footnote{\citeauthor{girvan2013-virtualworld} was contacted in regards to this technical document; she noted that an article was in press concerning the same content, but was not publicly available at the time of this writing.}
has also gathered properties for a virtual world through literature review and compiled a definition of a virtual world, namely ``a persistent, simulated and immersive environment, facilitated by networked computers, providing multiple users with avatars and communication tools with which to act and interact in-world and in real-time''. All properties are added in the study below.
 
\todoin{
~\cite{girvan201?-???}
 
WoW phasing, use against~\cite{bell2008-virtualworld}: Same virtual, different temporal ?frames?~gothman?
}

\citeN{schroeder2008-defining} argues for the definition of a virtual environment or virtual \mbox{reality} technology to be ``a computer-generated display that allows or compels the user (or users) to have a sense of being present in an environment other than the one they are actually in, and to interact with that environment''; the `being there' is then translated to a multi-user collaborative shared virtual environment, where participants can interact. Although sufficiently vague to perhaps remain valid into the future, the definition can not be used to determine an architecture for a virtual world using, for example, instances, cloud computing, MANets, \textit{etc}. 
\citeauthor{schroeder2008-defining} continues to differentiate virtual reality or virtual environments from virtual worlds on the basis that virtual worlds have been applied to persistent online social spaces. The definition by \citeauthor{schroeder2008-defining} does not discern if Facebook or the Internet at large is considered a virtual world. The properties of multi-user, sharing, interaction and online are examined in the study below.

According to \citeN{spence2008-demographics}, a virtual world is a ``persistent, synthetic, three dimensional, non-game centric space''; making a distinction between games and social spaces. No such distinction is made here, because (social) play can be considered part of gaming\todo{REF}. %
Applying his definition, \citeauthor{spence2008-demographics} was able to classify 37\% of the surveyed projects as virtual worlds, leaving 50\% to be classified somewhere on a continuum of hybrids. In addition, \citeauthor{spence2008-demographics} considers the Metaverse a virtual world, whereas in this article, the Metaverse is considered to be similar to the Internet, in accordance with~\cite{dionisio2013-metaverse}.

In literature, \citeN{singhal1999-netve} can often be found as an underlying reference with respect to a virtual world. 
Five common features are said to distinguish a net-VE: a shared senses of space, a shared sense of presence, a shared sense of time, a way to communicate, and a way to share~\cite[p.3]{singhal1999-netve}. Unfortunately, the features are described too vague to be useful for classifying technologies \eg a `shared sense of space' with respect to the instance dungeons of World of Warcraft or MANets; and, a `shared sense of time' with respect to multiple possible abstractions of time. It is unclear why a `way to communicate' and a `way to share' are chosen as features rather than action and interaction.
A graphics engine and display are mentioned as the cornerstones of a net-VE, and so propertiesÊrelated to these are added to the study. %

%% file: incl/4-approach.tex
\sectioncaps{Methodology and Scope}
\label{section:methodology}

Rather than obtain a definition through literature review, grounded theory~\cite{johannesson2014-designscience} (a systematic approach according to \citeN{creswell2013-qualitative}) is used \ie 
a definition can be formed from any possible criteria related to a virtual world, rather than being limited to definitions or criteria found in previous literature.
Sample technologies were chosen and analyzed for properties related to a virtual world. 
Initial sources of sample technologies and properties were: \citeN{girvan2013-virtualworld}, \citeN{bartle2003}, \citeN{bell2008-expanded}, and \citeN{combs2004-def}. 
Additional properties were sought, until the collection thereof, correctly classified all the sample technologies.
Theoretical sampling was used to select technologies that could pose a problem for the classification, until the point of `theoretical saturation' was reached; ``at this point, new empirical data does not help the researchers to further develop the theory''~\cite[p.48]{johannesson2014-designscience}. 
A total of 26 technologies (see the Appendix) %
were sampled before theoretical saturation was reached.
The most difficult properties to locate were those differentiating video games and virtual worlds, but thereafter sampling stopped because no other technology could be found in the consulted literature that was radically different enough to challenge the theory.

The novelty in using grounded theory to classify technologies that implement a virtual world, is that if a new technology emerges it can be handled \ie if the new technology challenges the current theory, properties can be added to the theory and the definition updated.

\subsection{Semantics}

Before presenting the study, it is important to clear up the semantics surrounding the term `virtual world'.
Perhaps one reason for the lack of definition, is because of a language problem. The word `virtual' can be used to mean ``being such in power, force, or effect, though not actually or expressly such''~\cite[`virtual']{dict.com2015}. This leads to an interpretation of, for example, a `virtual world of electronics' \ie there is almost an entire world of electronics or an imaginary world consisting entirely of only electronics. The meaning of `virtual' related to computers can be defined as ``temporarily simulated or extended by computer software''~\cite[`virtual']{dict.com2015}, leading to usages such as: a computer generated visualization of the digital world of electronics or a simulation of a world. The word `temporarily' in the definition can be justified by the fact that it is nearly impossible to create a truly persistent world~\cite{nevelsteen2016-persistence}. This article shall concentrate on the lattermost meaning of a virtual world, being a world simulated by computers, ruling out the imaginary. 
Note the distinction between: an (imaginary) world created in the minds of users while using a computer system (\eg perhaps while using a telephone) and a virtual environment that is simulated by a system, which then is perceived by users to be a world. %
In popular discourse, a virtual world is sometimes used to refer to the concept of a massive multiplayer online role-playing game (MMORPG), but a MMORPG is shown to be a higher level concept below. 
Also, in popular discourse, when dealing with `mixed reality'~\cite{milgram1999-mixed}, it is sometimes helpful to regard the system as ``a virtual world overlaid on the physical world''~\cite{lankoski2004-songsnorth} \ie the world of the virtual overlaid on the physical; those components that are virtual, in contrast to physical. 
And lastly, according to \citeN{schroeder2008-defining} ``the word `virtual' has come to mean anything online (as in `virtual money')''.
Speaking of a `virtual transfer' could be problematic in the sense of it being simulated by a computer \ie it didn't really happen; ergo, calling it an `electronic transfer' sidesteps the problem.
That `virtual money' is simulated is not a problem. \citeN{schroeder2008-defining} notes an ambiguity in the usage of `virtual money' to mean, money within a virtual world, which is a more specific type of simulated money. 
By definition, the term `online' means ``with or through a computer, especially over a network''~\cite[`online']{dict.com2015}, implying virtual and networking. The property of being online is added to the study below.

According to \citeN[p.26]{qvortrup2001-interaction}, a `world' is ``the all-encompassing context for the totality of human activities and experiences''. Creating such a world might prove impossible.
By creating a virtual world, an allegory of the physical world is modeled or simulated. Note, `world' is not equal to a planetary world \eg the science fiction universe of EVE Online~\cite{ccpgames2003-eve} is a virtual world which consist of an abundance of planetary worlds. A difficulty in obtaining a definition, as shall be seen, lies in quantifying the `world' of a potential virtual world \ie determining if the world qualifies as `worldly' enough\todo{perhaps why ppl use VE}. 
If a world is described as an environment that consists of people, places and things, sharing time and space, then respective properties can be added to the study, as a starting point.

\subsection{Virtual (Simulated) Environment (\VE)}
\label{section:ve}

For the purposes of this article, a \VE is considered to be minimally one that is: `wholly synthetic'~\cite{benford1998-boundaries}, simulated by a computing device and supports the least spatial property of containment~\cite{benford1998-boundaries}. 
\citeN{milgram1999-mixed} clarify that a \VE is one ``which must necessarily be completely modeled in order to be rendered''; contrary to the real (physical) environment over which ``the computer does not possess, or does not attribute meaning to, any information about the content''.

It should be noted that there are several abstractions of simulation possible within a computing device \eg (in general) the micro instructions of the CPU simulate instructions of the CPU, which in turn simulate assembly or higher level languages, which in turn simulate the operating system of a computer, which in turn can simulate the software engine simulating a \VE. If a technology can provide for one or more (\MANY) `data spaces'~\cite{nevelsteen2015-pervasivemoo}, then the loaded state in \MANY data spaces can possibly constitute a \VE. 
It is possible to have \MANY environments, nested in other environments (\eg multiple software running on the OS); a virtual world is a specialization of \MANY~{\VE}s. This logic is counter to the initial choice of words by \citeN{bartle2003}, who found `world' more encompassing, because it is ``not so easy to see how you could have several worlds within an environment''~\cite[\#9]{combs2004-def}. Of course, when Bartle was co-creating MUD~\cite{bartle2003}, EVE Online didn't exist \ie an abundance of planetary worlds.
In his updated definition, \citeN{bartle2010-history} specifies that a virtual world should be `automated', implementing a `set of rules (its physics)'. The property of a virtual world being automated is captured here by referring to a simulation, where the simulated \VE has its rules or physics.

\subsection{Architecture}
\label{section:architecture}

If an architecture aims to support a virtual world, it must implement the various properties of a virtual world (\eg persistence). Architectures are often chosen to satisfy additional requirements \eg mobility, decentralization or scalability. 
Restrictions on computing resources, such as memory, computation and networking, often determine how much of a virtual world can be loaded on a single client or server; some tactics for dividing up world space are discussed in Section~\ref{section:one_shard}. Scalability is often a reason to choose a particular architecture that is related to size (see Section~\ref{section:size}). Various communication architectures can be used to allow clients to connect to a virtual world (see Section~\ref{section:communication_architectures}).

\sectioncaps{Grounded Theory}
\label{section:grounded_theory} 

Each section below defines or clarifies a property that is used to classify sampled technologies (if possible) in the grounded theory;  classification details can be found in the Appendix in a underlying section with the same section heading. Final results are recorded in Table~\ref{table:results}. As a guide, abbreviations found in parenthesis of each section heading correspond to the column headers of Table~\ref{table:results}, Table~\ref{table:logical_and} and those found in the Appendix.

\subsection{Simulated (\ST)/Virtual Temporality (\VT)}

Similar to the simulation of a \VE, there are several layers of abstractions present when examining temporality in computers \eg the internal hardware might run at a particular frequency (a number of cycles per second, Hertz); computer programs are perhaps triggered according to an internal clock; and the operating system might have another abstraction of time \ie there can be an arbitrary number of abstractions of time. 

If the internal clock frequency is referred to as hardware temporality, the abstraction of time that is simulated using hardware temporality can be referred to as \ST; a computerized clock (usually at an accuracy of milli- or nanoseconds) that emulates `real-world'~\cite{zagal2007} time, but which can be altered independent of real-world time, %
since it is simulated.
Computers systems often maintain \ST as an `epoch'~\cite[System time]{wikipedia2016-org} %
plus a number of time units (an `offset').
Instrumenting real-world time incurs instrumentation error (\ie \ST is not exactly equal to real-world time), and so, real world events are recorded according to \ST as they enter a system~\cite{nevelsteenDRAFT-spatiotemporal}. %
If the value of \ST is recorded directly in the loaded state of software, the value represents `objective time'~\cite{nevelsteenDRAFT-spatiotemporal} (\eg \mbox{\texttt{Thu May 5 07:56:40 UTC+1 2016}}), %
whereas if a value relative to the \ST is recorded (\eg elapsed time), the value represents `subjective time'~\cite{nevelsteenDRAFT-spatiotemporal} (\eg an offset of 823 seconds, assuming the epoch of \ST). 

Software running in a computer (\eg a virtual world engine~\cite{nevelsteen2015-pervasivemoo}) will run according to \ST, but can simulate another abstraction of time, called \VT, for its loaded state (\eg a \VE). 
\VT can be simulated by establishing an epoch for \VT (\eg zero or \ST when the state was loaded) plus an offset. 
Since \VT can be altered independently of \ST, it is precisely \VT that allows a loaded state to be persisted to storage and reloaded (see Section~\ref{section:persistence}), or time in a \VE to be paused (see Section~\ref{section:non_pausable}), without side effects \eg by shifting the epoch of \VT or by pausing the increase of the offset, respectively.

\subsection{Shared Temporality (\OT)}
\label{section:shared_temporality}

Interacting agents, \HA or \SA (see Section~\ref{section:human_agents} and \ref{section:software_agents} respectively), can cumulatively share abstractions of time (their timelines overlap): real-world, hardware, \ST or \VT.
\OT means agents are able to share simultaneously real-world time and exactly one (\ONE) \VT, with other agents or entities they are acting upon. %
The following are examples of how abstractions of time are cumulatively shared and not:
\begin{compactitem}[$\circ$] %
\item two people in a telephone conversation share real-world time;
\item two players playing email chess, making moves at complete different times of the day (same timezone) do \textbf{not} share real-world time;
\item two applications running simultaneously on the same machine share real-world time, hardware temporality and \ST; if each application implements a \VT local to the application, applications do \textbf{not} share \VT; 
\item two machines running simultaneously in different timezones do \textbf{not} share hardware temporality, but can be synchronized to share \ST \eg a world time; 
\item two applications intermittently context switching do \textbf{not} share \ST;
\item two applications running simultaneously on different CPUs or hardware machines do \textbf{not} share hardware temporality; 
\item two applications time sharing \VT (\ie taking turns elapsing the virtual time) can do this by \textbf{not} sharing real-world time;
\item two agents in a \VE with \VT running on a single machine, share all abstractions of time simultaneously;
\item and, finally, 
two clients connected to a \VE distributed over \MANY machines do \textbf{not} share \ST if they are not synchronized, but can still share \VT. 
\end{compactitem}

\REQUIRES \VT

\newpage
\subsection{Real-time (\Rt)}
\label{section:real-time}

For a \VE to be considered \Rt (not turn-based and not tick-based, where agents must wait for other agents to complete their actions before a new round can begin~\cite{zagal2007}), it implies: 
\begin{compactitem}[$\circ$] %
\item agents can perform actions simultaneously~\cite{wikipedia2015-timekeeping} (the synchronous communication mentioned by \citeN{bell2008-expanded} is a subset of this interaction); 
\item and, the game world is available and action immediacy is part of the game design~\cite{zagal2007}.
\end{compactitem}
\smallskip
Note that \Rt is not the same as real-time computing constraints\todo{REF:\Rt computing constraints?}; \Rt is contrary to turn-based, whereas a \VE with real-time computing constraints must complete computations before certain time deadlines. 
Performance is an issue for the \Rt property, but missing deadlines not does constitute system failure as in real-time computing.
Although latency reduces the feeling of \Rt~\cite{zagal2007}, massive amounts of latency does not invalidate the \Rt property for a \VE, just a very slow one \eg massive space battles in EVE Online have been known to reduce graphics to a slideshow of images~\cite{drain2008-eve}.

\subsection{Virtual Spatiality (\VS)}

\citeN{aarseth2000-allegories} states that ``computer games are allegories of [physical] space''. This is similar to the feature, determined by \citeN{singhal1999-netve}, that \mbox{net-VEs} provide a `sense of space' for all participants. And, similar to what Raph Koster refers to as a ``spatially based depiction'' of a \VE~\cite[\#33]{combs2004-def}. 
\citeN{benford1998-boundaries} have classified a scale of spatiality, with `containment only' on one end of the spectrum and a `shared spatial frame' (\eg a Cartesian coordinate system) on the other, having identifyied ``fundamental physical spatial properties such as containment, topology, distance, orientation, and movement''. 
Movement is an important factor in \VS~\cite{benford1998-boundaries} \eg in MUD~\cite{trubshaw1978-mud} you move through seemingly continuous space~\cite[p.28]{qvortrup2001-interaction}, but in discrete steps; in a graphical world the equivalent would be moving from pixel to pixel in the frame of reference.

\subsection{Shared Spatiality (\OS)}

\citeN{singhal1999-netve} describe `a shared sense of space' such that: ``all participants have the illusion of being located in the same place, such as in the same room, building, or terrain. That shared place represents a common location within which actions and interactions can take place. The place may be real or fictional''. %
Note that the description does not exclude single player games (\ie technologies with just \ONE human agent and \SA), %
even though the net-VE, by \citeN{singhal1999-netve}, is described as a multiple user systems. The description is problematic for two reasons: First, telepresence, where users see into each other's physical space, is included in the set of net-VEs. If the definition herein is for a `pure' virtual world, then ``the place may be real'' should be excluded. Second, ``in the same place, such as in the same room, building, or terrain'' is too restrictive. If `the same place' (see Section~\ref{section:place}) can be understood as the same virtual world, with (possibly multiple) different terrain and physics (\eg it is entirely possible to have different planetary worlds within the same virtual world, such as in WoW), then this part of the description is ok.

\citeN{benford1998-boundaries} state that: ``Logically, the shared space in which a cooperative activity occurs can be defined to be those aspects of the system which are independent of any single participant or group of participants and which are agreed on by the participants''. This description is problematic, because the shared space of a virtual world is dependent on the architecture serving that world \eg~if \OS is to be ``independent of any single participant or group of participants'', then having a single participant or group of participants host the virtual world is not allowed. Property \P handles the persistence of a shared resource (see Section~\ref{section:persistence}).

\REQUIRES \VS and \Sh

\subsection{\ONE Shard (\Sh)}
\label{section:one_shard}

In an attempt to differentiate one world or many worlds, technologies can be examined to see if they are contained within \ONE shared data space, \ONE \textit{shard}\footnote{
``This term is an Ultima Online fiction to explain how come there are multiple copies of a supposedly single world''~\cite{bartle2003}.
}, rather than \MANY.
Two ways, to partition a \VS to allow for scalability, are regions and shards~\cite{liu2012-melding,dionisio2013-metaverse}. Using regions, the \VS is divided into static or dynamic areas, with each area handled by a different group of servers; players can still move throughout the entire \VS. With shards, players are divided up into groups and assigned to a unique copy (a shard) of the \VS, with each shard handled by a different group of servers; players are prohibited from moving between shards. Shards are copies of the same \VS that do not synchronize with each other.
Note, that a shard can be divided up into regions. 
To examine for the presence of \Sh, technologies must be examined to see if players share a common data space; with regions players still share \VS, with \MANY shards, they do not.

\subsection{Size}
\label{section:size}

Given \VS, the property of size is added to this study as possibly characterizing a virtual world \ie a \VE that is too small or too big (\eg a universe), can or can~not be a virtual world.
\citeN{bartle2003} states that, ``virtual worlds imbue a sense of size'' and that ``size is affected by many factors''. 
``If you can teleport anywhere, the world will feel smaller''~\cite[p.261]{bartle2003} \ie the `apparent size', the one perceived by the user, will be smaller. One way to judge `absolute size' is to calculate the number of discrete points in the environment and the speed of travel through the environment. But, both the apparent size of an environment and how big the absolute size needs to be (\ie a \VE is worldly enough) are subjective. So, size is not a determinant property of virtual worlds. \citeN[p.28]{qvortrup2001-interaction} states that virtual (3D) space has the general property of being ``geometrically finite, \ie it makes up a `bounded', `delimited' world'', but current computer systems are capable of simulating theoretically infinitely large virtual worlds \eg in Minecraft~\cite{quiring2015-placemaking}. The implementation of a \VE largely dictates how freely a player can move through \VS. The existence of teleporters invalidates the size metric further and does not differentiate between a world or multiple worlds. A world can be divided into `mini worlds' that have a free format and teleporters between them~\cite{yahyavi2013-p2pmmog}. 

\subsection{Indoor/Outdoor}

It can be questioned whether the indoor or outdoor nature of a \VE characterizes a virtual world \eg is both an open outdoor and an indoor space required for a \VE to be considered a virtual world?
\citeN{aarseth2000-allegories} specifies ``two different spatial representations: the open landscape, found mostly in the `simulation-oriented' games, and the closed labyrinths found in the adventure and action games''. \citeauthor{aarseth2000-allegories} calls the open landscape `outdoor', and the closed labyrinths as `indoor'. 

However, the concept of indoor or outdoor is based on the perception of the user and the cultural aspects of the game. \citeN{aarseth2000-allegories} states that even from a cultural perspective, ``what seems like an outdoors game is very much of the indoor variety: discontinuous, labyrinthine, full of carefully constructed obstacles''. From a technical perspective, there is also a problem, since the basic spatial property of a \VE is said to be containment; all entities are contained within the environment and would be deemed indoor. Even a simulated infinitely large world is still contained within a \VE. Indoor/outdoor is not a determinant property of virtual worlds.

\subsection{Places}
\label{section:place}

\citeN{singhal1999-netve} describe a `shared sense of space' as a `shared place' wherein interaction can happen. It is unclear if words of the description were chosen with regard to the `space versus place' debate~\cite{benford1998-boundaries}, or not. Considering the social meaning of `place', it is only possible to create a space which has the opportunity to adopt meaning and become a place \eg technology can create the space of a virtual world, but when it adopts meaning the virtual world becomes a place; the same is true for spaces within a virtual world.
Because place is socially determined, place is not a determinant property of virtual worlds.

\subsection{Things or Phenomena}

Simulating things or phenomena means implementing in software a game object\footnote{ 
``The collection of object types that make up a game is called the \textit{game object model}. The game object model provides a real-time simulation of a heterogeneous collection of objects in the virtual game world''~\cite[original italics]{gregory2014-v2}. 
}
with the most trivial needed properties (\eg position, orientation, description or visual representation) and deriving more advanced game objects by adding functionality or specialization to the derived game object. Even entities, such as the ground to walk on or water, might be implemented as a single or collection of game objects. For environmental effects such as lighting or heat, that have a source, but where only the effect is modeled, the effect is often attributed to the environment \eg the sun produces heat, but the sun is not modeled, only the effect of temperature on other game objects. A world will contain both ephemeral things, such as spoken words, and non-ephemeral things, such as rigid bodies. Supporting the non-ephemeral requires the world to support \P (see Section~\ref{section:persistence}). If \SA is supported, then the system is rich enough to support things and phenomena also.

\REQUIRES \SA

\subsection{\MANY Human Agents (\HA)}
\label{section:human_agents}

A \VE with exactly zero (\ZERO) humans can be equated to that of a pure simulation \eg a MUD where no players ever connect. Analyzing technologies for their support of one, two, three, four~\dots~and an increasing amount of humans leads to a subjective debate on how many humans is enough for a \VE to be considered a virtual world \eg \citeN{bartle2010-history} estimates that 100,000 players would be enough to constitute a massive world by todays standards, and \citeN{yahyavi2013-p2pmmog} state a massive game is capable of supporting hundreds or thousands of players.

\subsection{\MANY Software Agents (\SA)}
\label{section:software_agents}

Because a virtual world is a simulation, one can attempt to simulate \HA. 
\citeN[original italics]{velde1995-cogarch} states that: ``The concept of \textit{agent} refers to a system that can be differentiated from its environment and is capable of direct and continued interaction with that environment. Such a system `exists' in its environment and has with it an observable interaction, which is called its \textit{behavior}''. 
\citeN[pp.271-2]{shoham1997-aop} has defined a \textit{software agent} to be one ``that functions continuously and autonomously in an environment in which other processes take place and other agents exist. [\dots] The sense of `autonomy' is not precise, but the term is taken to mean that the agents' activities do not require constant human guidance or intervention''.
If \VT is supported\footnote{
It is possible to run agents off other temporalities (\eg \ST using objective or subjective time), but in order to be able to persist the state of the agent (see Section~\ref{section:persistence}) 
and maintain a pure virtual world (see Mixed Reality in Section~\ref{section:mixed_reality}), \VT is required.
}, 
agents that activate at a regular interval are considered running continuously. 
According to \citeN[p.39]{russell2010-ai}, an agent that lacks autonomy ``relies on the prior knowledge of its designer rather than on its own percepts [\dots] A rational agent should be autonomous -- it should learn what it can to compensate for partial or incorrect prior knowledge'' \ie autonomy assumes learning. The more relaxed definition of an autonomous agent, by \citeN{shoham1997-aop}, is adopted here which does not assume learning. 
According to \citeN[p.39]{russell2010-ai} the most simple agents possible are simple reflex agents, which ``select actions on the basis of the current percept, ignoring the rest of the percept history''.
It is through \SA, that if a virtual world has \ZERO users, the simulation will produce events \ie develop internally.

\REQUIRES \VT

\subsection{Virtual Interaction (\Ix)}
\label{section:virtual_interaction}

``The concept of `interaction' generally means `exchange', `interplay', or `mutual influence'~''~\cite[p.34]{qvortrup2001-interaction}. It is important to make the distinction between the exchange of interaction and the `to act upon' or `react to a stimulus' \eg users act upon user interface entities and a player might react to the stimulus of heat or gravity. In order for exchange to be possible, agents other than humans, must be capable of interaction \eg \SA. 
This property means that users can both: interact with people~\cite{girvan2013-virtualworld} (other \HA in the virtual world) and `interact' with the world~\cite{girvan2013-virtualworld,bartle2010-history} (interact with \SA and act/react to things and the environment). Both the `a way to communicate' and `a way to share', referred to by \citeN[p.3]{singhal1999-netve}, is not handled separately, but categorized here under \Ix.
If two virtual entities are to interact, they must have virtual proximity~\cite{qvortrup2001-interaction}, requiring the entities simultaneously share time (\OT) and space (\OS) \ie they must share simultaneously real-world time and \ONE VT.

\REQUIRES \SA, \OT, \OS and by consequence \Sh

\subsection{non-Pausable (\nZ)}
\label{section:non_pausable}

Considering the set of technologies that support \Rt, there is a subset that have an `active pause system' \cite{wikipedia2015-timekeeping} \ie if \VT is supported\footnote{
Those technologies that do not support at least one shared \VT (\texttt{V} or \texttt{V+} in Section~\ref{section:appendix_ot}) are automatically \nZ (unless not \Rt). A computer system usually only allows for the setting of \ST (the current time), but not pausing relative to real-world time \ie becoming unsynchronized. 
}%
, agents can be allowed to effectively freeze \VT relative to real-world time. Satisfying the \nZ property requires the \textbf{absence} of an active pause (or that its use is limited to operators only, for the purpose of maintenance). According to \citeN{bell2008-expanded} the \nZ property is part of \P (see Section~\ref{section:persistence}).

\REQUIRES \Rt

\subsection{Persistence (\P, world (\wP), data (\dP)) and Pseudo-}
\label{section:persistence}

Bartle states that a persistent virtual world ``continues to exist and develop internally even when there are no people interacting with it''~\cite{bartle2003}. The criterion that a virtual world should `develop internally' is handled by \SA (see Section~\ref{section:software_agents}), 
which constantly alter the world (\eg as in MUD1). Later \citeauthor{bartle2003} revised his persistence criterion, removing the need for internal development, stating: ``if you stop playing then come back later, the virtual world will have continued to exist in your absence''~\cite{bartle2010-history}.
\citeN{bell2008-expanded} use Bartle's original criterion, considering \P to differentiate video games and virtual worlds on the basis that, ``a virtual world cannot be paused''; pausing is handled by \nZ (see Section~\ref{section:non_pausable}). 
Bartle's criterion that the world must ``continues to exist'' can be referred to as `world persistence' (\wP)~\cite{nevelsteen2016-persistence} \ie the world continues to exist and is available to the players when they want to access it. To ensure world data is not lost in the event of system failure, the system must support `data persistence' (\dP)~\cite{nevelsteen2015-pervasivemoo,nevelsteen2016-persistence} for all data spaces that hold non-ephemeral data. It is precisely \VT which allows for a loaded state to be persisted to storage and reloaded without side effects; without the abstraction of time, that world's time would freeze during storage, suddenly rewind on a rollback or fast-forward to the current real-world time on load. 
A hybrid based on \ST and \VT is possible but then the world would be mixed reality (see Section~\ref{section:mixed_reality}).

The problem with persistence is that it can be simulated, without raising awareness to the player \eg if a world is made available to players at all times they want to access it, it is seemingly always in existence; a straightforward way to achieve this is to have scheduled play times around down times~\cite{nevelsteen2015-pervasivemoo}. 
The simulation of \P can be generalized as `pseudo-persistence'~\cite{nevelsteen2016-persistence}.
Taking pseudo-persistence into account, \citeauthor{bartle2003}'s original definition of \P is superior to that of \citeyear{bartle2010-history}, because a world can be kept in existence through player interaction, simulating \wP~\cite{nevelsteen2016-persistence}.
To simulate \dP, world data can be loaded and made available at specific times when players, relevant to that data, are playing~\cite{soderlund2009-proximity}. 
A straightforward way to simulate \dP is to avoid needing it \ie simply have no data to keep track of. 

Also related to virtual worlds, but not to be confused with \wP, is the concept of `game world persistence' \cite{nevelsteen2016-persistence} \ie events and activities in the game world which do not `reset'~\cite{keegan1997-classification,brennan2009-pesky} and are not interrupted by real world events (\eg a bathroom break). Because game world persistence is from a cultural perspective, the concept is beyond the scope of this article.

\subsection{Centricity}
\label{section:centricity}

Centricity is defined by \citeN{milgram1999-mixed} as ``the extent to which a human observer's viewpoint is removed from the `ownship', that is, from the nominal viewpoint with respect to the viewer's own avatar, or own vehicle, or own manipulator within the task space''. The concept of centricity is generalized into a continuum between `egocentric' and `exocentric'~\cite{milgram1999-mixed}, with illustrative intermediate points being: ego-referenced, tethered or world referenced. In a 3D world, a 1st-person perspective can be created using an ego-referenced viewpoint and a 3rd-person perspective through either a 3D tethered (sometimes referred to as `over the shoulder') or a 3D world referenced viewpoint. 
\citeN[p.30]{qvortrup2001-interaction} defines two avatar categories: ``user-in-avatar, which is a representation of the human user in the virtual 3D world and the designer-in-avatar, which is a representation of the designer, developer or creator of the 3D world (sometimes called `God')''. These two categories can be enabled using various grades of centricity; the ego-referenced and tethered viewpoints can be used for the user-in-avatar representation, whereas world-referenced viewpoints can be used for the designer-in-avatar representation. 

Centricity is difficult to apply to a technology that is not 3D (\eg text-based interfaces) and many technologies offer multiple viewpoints with option to switch between them (\eg in WoW); such a change does not effect the worldliness of the \VE or whether an avatar is present (see Section~\ref{section:avatar}). Centricity is not a determinant property of virtual worlds, but a way to visualize them (see Section~\ref{section:visualization}).

\subsection{Mediation}

A desktop class computer has been typically used in the past to access virtual worlds. As computing capacity rises in handheld devices (\eg tablets and smartphones), they too are being used to access virtual worlds. How a virtual world is accessed (\ie the mediating technology), does not and should not determine its worldliness \eg accessing a text-based MUD on a modern smartphone does not effect the MUD. A virtual world might be accessed simultaneously by the same user through different devices and each device might offer a different functionality. Additionally, who knows what technology will be used to access future virtual worlds \eg head-mounted displays.

\subsection{Visual Representation}
\label{section:visualization}

It is possible to have multiple representations of the same world simultaneously on different different devices \ie support for crossmedia. This means that the visual representations of the game (\eg text, 2D, 2.5D and 3D) is not a determinant property of virtual worlds.

\subsection{Communication Architecture}
\label{section:communication_architectures}

Various communication architectures can be used to allow clients to connect to a virtual world \eg local on computer or device; client connecting to single server or a cluster of servers; peer-to-peer systems with or without additional infrastructure; or, many clients connecting to a cloud-computing platform~\cite{yahyavi2013-p2pmmog}. A particular communication architecture is often chosen to gain scalability or maintain control over the world~\cite{yahyavi2013-p2pmmog}. Although networking is required to support \HA, there are virtual worlds that allow for both local and network access (\eg MUD~\cite{bartle2010-history}). The property of communication architecture does not classify virtual worlds, because it is widely used by diverse communication tools, games and web services \ie communication architecture is not a determinant property of virtual worlds.

\subsection{Online}

When virtual worlds are discussed, the property of the \VE being online is sometimes brought into the discussion~\cite{schroeder2008-defining}. As per the definition~\cite[online]{dict.com2015}, accessing something `online' means ``with or through a computer, especially over a network''. This property is therefore broken down into \HA (see Section~\ref{section:human_agents}) and Communication Architecture (see Section~\ref{section:communication_architectures}).

\subsection{\MANY Avatar (\Av)}
\label{section:avatar}

Described as ``extensions of ourselves''~\cite[p.26]{qvortrup2001-interaction}, there are many definitions of the term `avatar'.
\citeN{bartle2010-history} describes an avatar as a `virtual self' where ``each player identifies with a unique entity within the virtual world (their character) through which all their in-world activity is channelled''. 
According to \citeN{bell2008-expanded}, ``any digital representation (graphical or textual), beyond a simple label or name, that has agency (an ability to perform actions), presence, and is controlled by a human agent in real time is an avatar''. %
The real-time aspect is handled by \Rt.
There is ambiguity in the situation where a player, in a single user environment, has an ego-referenced viewpoint \ie they can not look at themselves (\eg in Zork~\cite{supnik1994-zork} or in 1st-person perspective in a 3D environment); if no digital representation is present, does the player actually have one? Even without an avatar, \SA must be able to react to the presence of the player \ie some entity. In this regard, \citeauthor{bartle2010-history}'s definition of the virtual self is superior to that of \citeN{bell2008-expanded}; since the virtual self is defined as an entity through which presence can be detected and interacted with, rather than a graphical or textual representation. 

Although it is often true that virtual worlds assign a unique entity to the player, there is evidence that this is not a requirement. For example, in WoW, there are users that purchase multiple accounts and control multiple avatars simultaneously through multiple installations of the game; a technique that has been referred to as `multiboxing'~\cite{wowwiki2015-multiboxing}. Because multiboxing can be used to override the unique avatar feature of WoW, the definition by \citeN{bartle2010-history} shall be used here, but without the uniqueness criterion \ie \MANY avatar.

\subsection{Immersion}

Immersion has been researched in different domains \eg virtual reality, game research and interface design~\cite{brown2004-grounded}. Immersion is a qualitative property tied to the perception of the user, so it would seem that creators of virtual worlds can only aim for an immersive experience \ie immersion is not a determinant property. 

\citeN{benford1998-boundaries} introduce theÊdimension of `transportation' (which they compare with immersion), which ``concerns the extent to which a group of participants and objects leave behind their local space and enter into some new remote space in order to meet with others, versus the extent to which they remain in their local space and the remote participants and objects are brought to them''. 
With respect to defining a virtual world, the dimension is problematic \eg augmented reality, projected collaborative {\VE}s and immersive collaborative {\VE}s are on very different parts of the transportation continuum, but each of those {\VE}s could be a virtual world. The extent to which a user leaves their local space behind, does not determine if the \VE is a world.

\todoin{
"All but one game mentioned as totally immersive was a first person perspective game. Also role-playing games were mentioned, where the gamers assumed a character."~\cite{brown2004-grounded}.
}

\subsection{Presence}

A sense of presence is sometimes used interchangeably with immersion~\cite{coquillart2011-vr}, remarked as being required for immersion~\cite{brown2004-grounded} or \textit{vice versa}~\cite{slater1997-five}. 
Like immersion, presence is also a qualitative property tied to the perception of the user \ie presence is also not a determinant property for virtual worlds.

%% file: incl/results.tex
%!TEX root = ../biblatex-plgd/biblatex-plgd-virtualworld.tex

%:--- TABLE, Part 1

% http://tex.stackexchange.com/questions/83860/remove-page-number-from-just-one-float-page
\afterpage{%
\thispagestyle{plain}

\begin{table}[h!]
% can also use \centering instead, removes empty line between caption and table
%\begin{center}

%\centering
% http://tex.stackexchange.com/questions/163246/resize-a-tabular-object-to-textwidth
%\begin{adjustbox}{max width=\textwidth}
\begin{adjustbox}{scale=0.8}

\ttfamily\selectfont
\begin{tabular}{l|cl|c|c|c|c|l|c|cl|c|}

\textsc{technology}		& \OT
								&
												& \Rt
														& \OS    				
																& \Sh   		
																		& \SA
																				& \Ix		
																						& \nZ
																								& \P
																										&
																																& \Av
																																		\\
\hline                                                                  
email chess				& \R	& \EQ RS		& \N	& \Y 	& \Y	& \R 	& \R	& \R	& \Y	& \EQ \wP\AND \dP		& \N	\\ 
calculator				& \R	& \EQ RH		& \Y	& \R	& \Y	& \R 	& \R	& \Y	& \N	&						& \N	\\ 
CAD						& \Y	& \EQ RHS[V]	& \Y	& \Y	& \Y	& \N 	& \R	& \N	& \N 	& \EQ \wP\XOR \dP		& \N	\\ 
Google Docs				& \R	& \EQ RS		& \Y	& \R	& \Y	& \R 	& \R	& \Y	& \Y	& \EQ \wP\AND \dP		& \Y	\\
\hline                                                                                                                      
Skype incl.~video	 	& \R	& \EQ RS		& \Y	& \R	& \Y	& \R 	& \R	& \Y	& \Y	& \EQ \wP\AND \dP		& \Y	\\
IRC						& \R	& \EQ R[S]		& \Y	& \R	& \N	& \R 	& \R	& \Y	& \Y	& \EQ \wP\AND \dP		& \Y	\\
\hline                                                                                                                      
Zork					& \R	& \EQ RH[S]		& \Y	& \Y	& \Y	& \R 	& \R	& \Y	& \N	& \EQ \wP\XOR \dP		& \Y	\\
Civ5					& \Y	& \EQ RS[HV]	& \N	& \Y	& \Y	& \Y	& \Y	& \R	& \N	& \EQ \wP\XOR \dP		& \Y	\\
Doom (Deathmatch) 		& \Y	& \EQ RHSV(RSV)	& \Y	& \Y	& \Y	& \Y	& \Y	& \N	& \N	& \EQ \wP\XOR \dP		& \Y	\\
League of Legends		& \Y	& \EQ RSV		& \Y	& \Y	& \Y	& \Y	& \Y	& \Y	& \N	& \EQ pseudo-\wP\AND dP	& \Y	\\
\hline                                                                                                                      
MUD 					& \Y	& \EQ R[H]SV	& \Y	& \Y	& \Y	& \Y	& \Y	& \Y	& \Y	& \EQ \wP\AND \dP		& \Y	\\
WoW, \ONE shard 		& \Y	& \EQ RSV		& \Y	& \Y	& \Y	& \Y	& \Y	& \Y	& \Y	& \EQ \wP\AND \dP		& \Y	\\
WoW, all shards 		& \N	& \EQ RS[V+]	& \Y	& \R	& \N	& \Y	& \R	& \Y	& \Y	& \EQ \wP\AND \dP		& \Y	\\
\hline                                                                                                                      
Facebook				& \R	& \EQ RS		& \Y	& \R	& \Y	& \R 	& \R	& \Y	& \Y	& \EQ \wP\AND \dP		& \Y	\\
the Internet			& \N	& \EQ R[SV+]	& \N	& \R	& \N	& \Y	& \R	& \R	& \Y	& \EQ \wP\AND \dP		& \Y	\\
\hline

\end{tabular}

\end{adjustbox}

\caption{{\normalfont Do the technologies support the properties of: Shared Temporality (\OT, \MANY (\texttt{+}) shared real-world (\texttt{R}), hardware (\texttt{H}), simulated (\texttt{S}) or virtual (\texttt{V}) time); Real-time (\Rt); Shared Spatiality (\OS); \ONE Shard (\Sh); \MANY Software Agents (\SA); Virtual Interaction (\Ix); non-Pausable (\nZ); Persistence (\P, world persistence (\wP) and/xor data persistence (\dP)); and an Avatar (\Av)? Yes (\Y), No (\N), or an underlying requirement is lacking (\R) (see the Appendix for details).}
}

\label{table:results}

%\end{center}
\end{table}%

\begin{table}[h!]
% can also use \centering instead, removes empty line between caption and table
%\begin{center}

%\centering
% http://tex.stackexchange.com/questions/163246/resize-a-tabular-object-to-textwidth
\begin{adjustbox}{scale=0.8}

\ttfamily\selectfont
\begin{tabular}{l|cl|}

\textsc{technology}		
						& $\forall$\texttt{p}:\AND
								& 													\\
						
\hline                  
email chess				& \N	& \EQ \VT[\OT,\SA,\Ix]\AND \Rt[\nZ]\AND \Av			\\ 
calculator				& \N	& \EQ \VT[\OT,\SA,\Ix]\AND \VS[\OS]\AND \P\AND \Av	\\ 
CAD						& \N	& \EQ \SA[\Ix]\AND \nZ\AND \P\AND \Av				\\ 
Google Docs				& \N	& \EQ \VT[\OT,\SA,\Ix]\AND \VS[\OS]					\\
\hline                  
Skype incl.~video	 	& \N	& \EQ \VT[\OT,\SA,\Ix]\AND \VS[\OS]					\\
IRC						& \N	& \EQ \VT[\OT,\SA,\Ix]\AND \VS[\OS]\AND \Sh			\\
\hline                  
Zork					& \N	& \EQ \VT[\OT,\SA,\Ix]\AND \P						\\
Civ5					& \N	& \EQ \Rt[\nZ]\AND \P								\\
Doom (Deathmatch) 		& \N	& \EQ \nZ\AND \P									\\
League of Legends		& \N	& \EQ \P											\\
\hline                  
MUD 					& \Y	&													\\
WoW, \ONE shard 		& \Y	&													\\
WoW, all shards 		& \N	& \EQ \OT\AND \Sh[\OS,\Ix]							\\
\hline                  
Facebook				& \N	& \EQ \VT[\OT,\SA,\Ix]\AND \VS[\OS]					\\
the Internet			& \N	& \EQ \OT\AND \Rt[\nZ]\AND \Sh[\OS,\Ix]				\\
\hline

\end{tabular}

\end{adjustbox}

\caption{{\normalfont The logical \AND~of all properties (values in brackets next to a property, means those properties fail due to lacking an underlying requirement; the properties Virtual Temporality (\VT) and Virtual Spatiality (\VS) can be found in the Appendix, see Table~\ref{table:sharedtime} and  Table~\ref{table:sharedspace} respectively).}
}

\label{table:logical_and}

%\end{center}
\end{table}%

\clearpage
}

%% file: incl/5-results.tex
\sectioncaps{Analysis and Result}

The aim of this work is not to evaluate how virtual-like or world-like a virtual world is, but to find a definition that delineates what is and is not a virtual world \eg a 3D virtual world with multiple players, a vast many places to visit, totally immersing each player might feel much more worldly than a MUD, but the definition should not exclude a MUD from being a virtual world.

\subsection{The Logical \AND~of Properties \texttt{($\forall$\texttt{p}:$\land$)}}

Having examined individual technologies (see the Appendix) for properties pertaining to virtual worlds (that are found in the subsections of Section~\ref{section:grounded_theory}), those properties are collected that together classify well known virtual worlds \eg MUD, localhost or with networking, and WoW. 
In this section, the resulting properties (see \mbox{Tables~\ref{table:results} and \ref{table:logical_and}}) of \OT, \Rt, \OS, \Sh, \SA, \Ix, \nZ, \P and \Av are assembled into a definition and the outcome justified. 

\subsubsection{Real-time Shared Spatiotemporal \VE}
\label{section:shared_spatiotemporal}
\label{section:results_interaction}

\VT ensures the world has its own time, independent from other times. A (not purely) virtual world can be built based on \ST (without \VT), but all events related to \ST are subjected to being rewound or fast-forwarded, in the case of rollbacks or downtime, respectively. \VT works together with \P, ensuring that events persisted to disk are independent of \ST \ie \VT can be translated along the timeline relative to real-world time and \ST, without effecting time-based in-game events. The \Rt property excludes turn-based, tick-based, and any other technologies that do not support agents simultaneously interacting with the world.
\OT combined with \OS ensures that the \VE conforms to a spatial representation with the likeness of a world and that agents share that world.

\subsubsection{Shared Spatiality within \ONE Shard}
\label{section:shared_spatiality_within_one_shard}

If a technology has more than one data space, \Sh dictates that those spaces must be merged at one point or another for them to be considered one virtual world; each other copy/shard is considered another world. The question at hand is how to deal with instances \ie are they part of the virtual world or not? A solution, is to judge each shard on its own merits as to whether it classifies as a virtual world or not. An instance can be considered a temporary shard (a copy of space to which a group of players is assigned), but not a virtual world, if it does not support the required properties \eg \P.
In Diablo~\cite{blizzard1996-diablo}, the virtual lobby is not a virtual world, because it lacks properties such as \VS. And, the world instances are not virtual worlds either. If the world instance would constantly synchronize with the virtual lobby, the lobby would provide \P for the instance.
It is specifically the \Sh property that allows for a virtual world to be divided in a distributed system, but still remain one world.
\OS and \Sh are combined to determine if technologies have a shared spatiality captured within \ONE shard. 

\subsubsection{Interacting Agents and Things}
\label{section:agentsor}

As discussed in Section~\ref{section:human_agents}, most technologies (\eg except a pure simulation of a world) support \HA. \citeauthor{bartle2003}'s criteria states that a persistent virtual world ``continues to exist and develop internally even when there are no people interacting with it''. From this it can be ascertained that the world must support \ZERO humans \ie a world that suddenly has no people connected to it doesn't loose its worldliness. This would then also allow for a virtual world to be created for dogs or other non-humans. If a world is to continue to develop internally with \ZERO humans, the world must support entities that can take action \ie support for \SA determines if there is enough potential intelligence in the system to cause sufficient worldly change.  
Combined with \Ix and the shared spatiotemporal \VE from above, this means humans can potentially virtually interact, in the same time and space, with other \HA or \SA, or act/react to things and the environment.
A technology that just supports \HA is a communication technology, but not necessarily a virtual world.

\subsubsection{Persistent Virtual World}
\label{section:result_persistence}
\label{section:result_notpausable}

\P ensures the world is available to \HA when they want to access it and where non-ephemeral changes in the world are preserved. \P combined with the shared spatiotemporal \VE from above ensures that the world is not a fleeting transient event. To satisfy the criteria by \citeN{bartle2003}, \P is combined with \SA. To satisfy the criteria by \citeN{bell2008-expanded}, \P is combined with \nZ, a property that seems to effectively segregate virtual worlds from video games, when combined with \P.
The result of these combinations is a real-time environment that is non-pausable where agents produce lasting change.

\subsubsection{Virtual Self as Avatar}
\label{section:virtualself_avatar}

\citeN{singhal1999-netve} define a net-VE as a multiuser environment, with each user represented as an avatar; users need a representation through which other users can detect presence and interact. Although it might simplify the definition, no assumption is made here that a virtual world is required to be multiuser, so a single user environment is still acceptable. 
The uniqueness criterion for a virtual self has been dismissed on the grounds that multiboxing does not alter the worldliness of a virtual world; \Av is the virtual entity which \citeauthor{bartle2010-history} has defined as a `virtual self', with the  uniqueness constraint dropped, so that agents (\HA or \SA) have something to interact with. If \MANY avatars are present, at least one avatar must be present in the shared spatiotemporal \VE from above.

\subsection{Definition: Virtual World}
\label{section:the_definition}

Having combined the determinant properties, it is now possible to state the definition for a virtual world (VW) as:

\theoremstyle{definition}
\begin{definition}%
\label{def:vw}
	A simulated environment where:
	\MANY agents can virtually interact with each other, act and react to things, phenomena and the environment;
	agents can be \ZERO or \MANY human(s), each represented by \MANY\footnote{keep in mind that \MANY means `one or more' \ie a virtual self is not required to be unique.} \mbox{entities} called a \mbox{`virtual self'} (an avatar), 
	or \MANY software agents;
	all action/reaction/interaction must happen in a real-time shared spatiotemporal non-pausable virtual environment; 
	the environment may consist of many data spaces, but the collection of data spaces should constitute a shared data space, 
	\ONE persistent shard.
\end{definition}

Properties not used in Definition~\ref{def:vw} can instead be use to distinguish between different virtual worlds, rather than classify a virtual world. 
\citeN{girvan2013-virtualworld} offers the properties: ``3D, educational, goal orientated, graphical and user-generated content''. 
 `3D' and `graphical' can be generalized to the visual representation of Section~\ref{section:visualization} and the following unused properties from the analysis in Section~\ref{section:grounded_theory} can also be used: size, indoor/outdoor, centricity, mediation, communication architecture, immersive or various degrees of presence. 
Some additional properties not mentioned in the analysis are: number of players, theme (\eg fantasy, science fiction), strategy type (\eg RPG, PvE, PvP) and in-game communication possibilities (\eg chat, VoIP).

%% file: incl/6-verification.tex
\graphicspath{{img/}} % specifies where the figures are stored

\sectioncaps{Verification}

In this section, Definition~\ref{def:vw} is first verified against the prominent existing definitions mentioned in related work. Thereafter, a form of `discriminant sampling'~\cite{creswell2013-qualitative} is used \ie selecting advanced contemporary technologies that were not in the study to see if the theory holds true for these additional samples, classifying advanced technologies, such as: a pseudo-persistent video game, a MANet, virtual reality, mixed reality, and the Metaverse.
The effectiveness of the obtained definition is demonstrated by creating an ontology of virtual worlds \ie clarifying acronyms commonly found in research and popular discourse
  
\subsection{Definition Comparisons}

A virtual world can be classified as a net-VE~\cite{singhal1999-netve}, where: the multiuser restriction is relaxed to allow local play and simulations with no humans;
the environment is required to be persistent. %
A `shared sense of space' and a `shared sense of time' can be assumed to be equal to the shared real-time spatiotemporal \VE in Definition~\ref{def:vw}, except for one problematic sentence in the definition. \citeauthor{singhal1999-netve} state that the place in a `shared sense of space', `may be real or fictional'. It is assumed that this statement is mentioned to include the physical aspect of a video conference, but not meant to include `pervasive systems'~\cite{nevelsteen2015-pervasivemoo}. If the latter is the case, there is not enough clarification provided in the rest of the definition to handle pervasiveness (see Section~\ref{section:mixed_reality}) \eg 
a virtual persona projected into the physical world, would seem to be included in definition, but a human interacting with that projection in the physical shared space, fails the avatar requirement by \citeauthor{singhal1999-netve}.
A `shared sense of time' by \citeauthor{singhal1999-netve} implies real-time interaction, but in Definition~\ref{def:vw} it is made explicit.
A `shared sense of presence' implies support for avatars that can be `synthetic entities'; again this aspect is made explicit in Definition~\ref{def:vw} through requiring software agents.
The need for an avatar, mentioned in the `shared sense of presence', is a consequence of a net-VE being defined as a multiuser environment. It is not clear if the avatar requirement has a uniqueness constraint.
A `way to communicate' and a `way to share' are considered part of action and interaction in Definition~\ref{def:vw}, with the different types of in-game communication left as not determinant of a virtual world \ie a distinguishing characteristic of various virtual worlds.
Using the two definitions, Google Docs~\cite{google2007-docs} classifies as a net-VE, but not as a virtual world.

\citeN{bartle2010-history} updated his definition of a virtual world and the persistence criterion, stating a virtual world to be ``an automated, shared, persistent environment with and through which people can interact in real time by means of a virtual self''. The automated property specifies a set of rules by which players can change the world, but doesn't mention how the world can develop internally (previously a part of \P~\cite{bartle2003}). The definition refers to a shared environment, where ``more than one player can be in the exact same virtual world at once'', but usage of the term `virtual world' in the description leads to a recursion; it can be assumed to mean a shared spatiotemporal \VE, but if the description is to serve as a definition this is problematic.
The definition mentions interaction within the virtual world for people, but does not mention virtual interaction in the world \eg  software agents.
\citeN{bartle2010-history} altered the persistence criterion to state: ``if you stop playing [and] then come back later, the virtual world will have continued to exist in your absence'', but \citeN{bartle2003}'s original definition of P is superior to that of \citeyear{bartle2010-history} (see Section~\ref{section:persistence}). 
The persistence criterion can be assumed to include the non-pausable criterion, by \citeauthor{bell2008-expanded}, but fails to account for the different between persistence in a single player video game and a virtual world.
\citeauthor{bartle2010-history}'s definition of a `virtual self', has been adopted in Definition~\ref{def:vw}, but the uniqueness criterion has been dropped.
Since time and space are not elaborated on, the definition by \citeauthor{bartle2010-history} can be said to describe a persistent communication technology \ie including virtual worlds, but also including technologies such as Google Docs.

Through literature review, \citeN{bell2008-expanded} offer the definition of ``a synchronous, persistent network of people, represented as avatars, facilitated by networked computers''. The term `synchronous' in the definition refers to synchronous communication. Although this seems to be similar to \Rt, synchronous is only applied to communication in their definition; \Rt requires real-time interaction, which includes communication. The term `synchronous' is also ambiguous when referring to communication \eg \citeN{bell2008-expanded} state ``a synchronous environment does not require the sender of a message to wait for the other party'', whereas an synchronous communication protocol is one where the sender \textbf{does} wait for a response. 
\citeauthor{bell2008-expanded} offer a citation for synchronous communication, but the underlying reference simply ensures that users are logged-on simultaneously, which is irrelevant in determining real-time interaction when both users are indeed logged on. 
Synchronous communication is said to imply the concept of continuous common time (similar to \OT), and bring about a sense of presence. 
\citeauthor{bell2008-expanded} use Bartle's %
criterion for persistence from \citeyear{bartle2003}, but add that ``a virtual world cannot be paused'', differentiating them from video games. This criterion has been adopted in Definition~\ref{def:vw}. 
A `persistent network of people' is mentioned in their definition, but persistence and a network of people is handled separately in their text; a persistent network of people is not the same as a world that stays in existence, unless worldliness is only defined in function of people. 
Linked to synchronous communication, \citeauthor{bell2008-expanded} refer to social grouping behavior and a primitive form of ecosystem, leaving a `network of people' to mean the social construct. In Definition~\ref{def:vw}, single user environments are allowed, not requiring a form of social networking, although social networking might be probable in a multiuser virtual world. 
\citeauthor{bell2008-expanded} define an avatar and state that it should have agency (\eg ``a Facebook profile does not have agency beyond its creator''), functioning ``like user-controlled puppets. Users command the actions of the avatar, but it is the avatar itself which performs the action''. This view would seem to deny an immersive ego-referenced perspective that virtual reality strives for (the reader is referred to~\cite{waggoner2009-myavatar} for a discussion on this matter).
The definition of a virtual self, by \citeN{bartle2010-history}, was adopted in Definition~\ref{def:vw}, because it provides an entity for other agents to interact with. 
The criterion `facilitated by networked computers' serves to ensure an environment simulated by computers is being referred to and that those computers are networked. Definition~\ref{def:vw} does not require multiple users or networking, allowing for virtual worlds that are pure simulations.

\citeN{girvan2013-virtualworld} states a virtual world to be: ``a persistent, simulated and immersive environment, facilitated by networked computers, providing multiple users with avatars and communication tools with which to act and interact in-world and in real-time''. The definition is compiled through literature review, but the sampling is heavily biased towards Second Life~\cite{spence2008-demographics}, with 47 out of 68 surveyed articles referring to Second Life. 
Definition~\ref{def:vw} agrees with many of the properties in this definition, but: `immersive' has been identified as a qualitative property that is subjective; as in the case of a net-VE above, Definition~\ref{def:vw} does not require multiple users or networking; and, \citeauthor{girvan2013-virtualworld} does not define shared time or space, which is especially problematic when dealing with video games, instances or ad-hoc networking. Definition~\ref{def:vw} is similar to the one from \citeauthor{girvan2013-virtualworld}, in the sense that action and interaction are handled individually. 

\begin{figure}[h] %
\includegraphics[width=\textwidth,height=200pt]{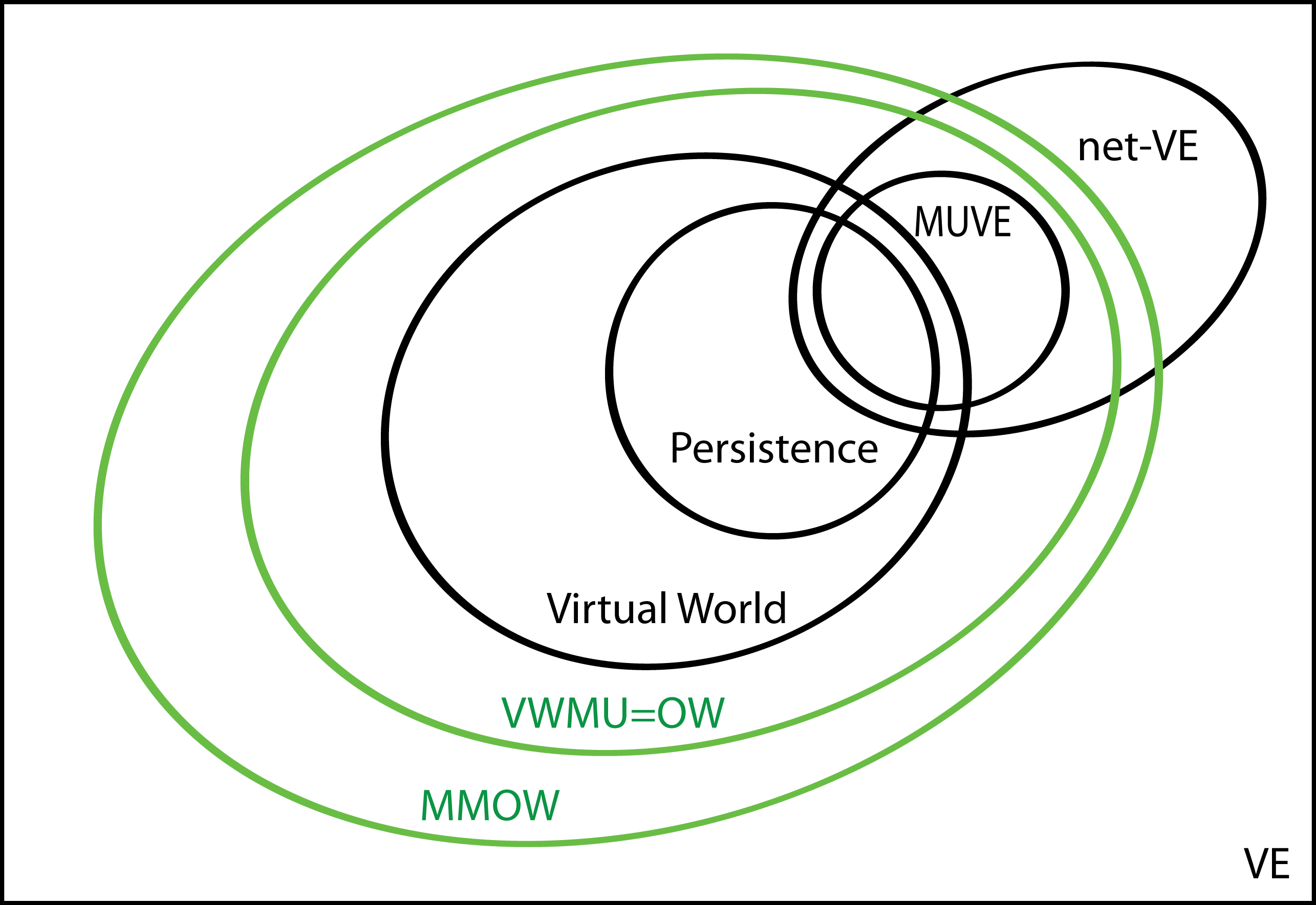}
\caption{Ontology of virtual worlds in relation to net-VEs and other sets}
\label{figure:ontology-vw}
\end{figure}

\subsection{Ontology of \VE}

If the net-VE definition by \citeN{singhal1999-netve} is used (assuming that 
multiple users (MU) implies networking) and Definition~\ref{def:vw} is used for a virtual world, then:

\begin{compactitem}[$\circ$]
\item a \VE that supports MU is then a MUVE, and equal to a subset of net-VEs;
\item there is a subset of MUVEs and net-VEs that support persistence~\cite{girvan2013-virtualworld};
\item the subset of net-VEs that support persistence, is a subset of the virtual worlds (VW) (according to the Definition~\ref{def:vw}) that support MU; %
\item since online implies both virtual and networking, a VW that supports MU is equal to an online world (OW); %
\item an OW supporting massive multiplayers (MM) is an MMOW; %
\item a \VE supporting MM is then a MMVE; %
\item an MMO that is a game is an MMOG~\cite{gregory2014-v2}, of which some are MMOWs~\cite{spence2008-demographics};  %
\item and finally, a MMOG that is of the role-playing type is an MMORPG, with the RPG type implying a MMOW that can be role-played in~\cite{gregory2014-v2}. %
\end{compactitem}

\bigskip

\begin{wrapfigure}{r}{0.5\textwidth}
  \vspace{-25pt}
  \begin{center}
	\includegraphics[width=\textwidth/2]{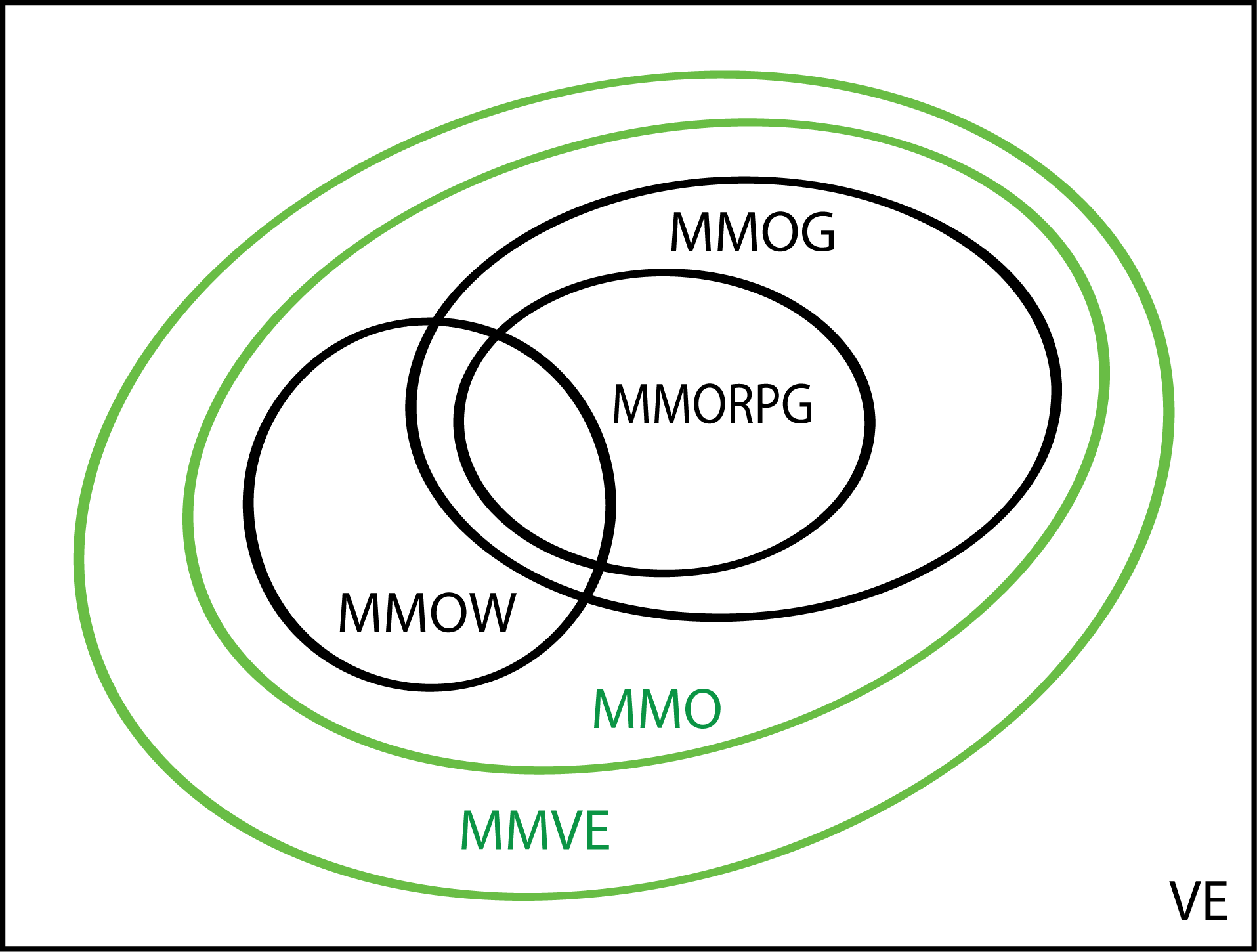}
  
  \end{center}
  \vspace{-8pt}
  \caption{Higher level ontological view of MMO's and the set of MMVE}
  \label{figure:ontology-mmve}
\end{wrapfigure}

The ontology for the acronyms is displayed in Figure~\ref{figure:ontology-vw} and~\ref{figure:ontology-mmve} (in no way does the size of the drawing imply the number of members in the set). The set of net-VEs contains the properties of: \OT, \Rt, \OS, \SA, \Ix, \Av, MU, and networking; the set for virtual worlds contains: \OT, \Rt, \OS, \Sh, \SA, \Ix, \nZ, \P and \Av; meaning a discrepancy on: \Sh, \nZ, \P, MU and networking. The set of VWMU or OW contains the additional properties of MU and networking; all sets starting with `MM' have the massive multiplayer property; and the general \VE has the properties of being simulated, spatial containment, and \MANY data spaces.

\bigskip
As seen in \cite{girvan2013-virtualworld}, these acronyms can be further coded with optional qualifiers (see Section~\ref{section:the_definition}) to create more subsets \eg in the case of a virtual inhabited three-dimensional worlds (VI3DWs)~\cite[p.25]{qvortrup2001-interaction} or an immersive virtual world (IVW)~\cite{girvan2013-virtualworld}. Interestingly about the latter example, is that if all virtual worlds were by definition immersive as in \cite{girvan2013-virtualworld}, the term `immersive' would be redundant in the acronym. %

\subsection{Evaluation Using Discriminant Sampling}

In this section, to further evaluate Definition~\ref{def:vw} and the underlying determinant properties, `discriminant sampling'~\cite{creswell2013-qualitative} is used \ie selecting advanced contemporary technologies that were not in the study to see if the theory holds true for these additional samples. But, first a clarification on the nuances that separate MUD and IRC.

\subsubsection{MUD and IRC}

A simulation of a world, is the only technology that fails to support \HA, and therefore could be said to fail for \VT \ie if the allegory of physical space of MUD is created through narrative and there are \ZERO humans to understand the narrative, then the allegory should be lost. WoW simulates an allegory of space, which is not based on narrative; if all players disconnect, software agents must still perceive an allegory of space. MUD has software agents which minimally understand the narrative (\eg `go south'), and therefore still perceive an allegory of space, leaving a MUD with \ZERO players to satisfy \VT. 

If scripting on IRC is included in the discussion, it is difficult to distinguish the technologies of MUD and IRC. IRC fails to have a shared spatiotemporal \VE; failing \SA and \Ix as a consequence. An IRC `bot' on a single IRC network, can be created to make use of \ST, so this bot can scripted to be a continuous autonomous software agent. A user of IRC can change `channels' with commands (similar to MUD) and each channel has a welcome statement (similar to a room description in MUD). Taking all this into account, the IRC network would satisfy all the properties, except one, to be considered a virtual world (albeit an empty one, since IRC has next to no world or player data that needs to be persisted). As soon as IRC was made more worldly, \P would become critical for persisting \SA and things that agents carry around or are lying around.

\subsubsection{Destiny}

From the analysis, it is apparent that it is difficult to differentiate between video games and virtual worlds. The only properties that seem to differentiate them are \nZ and \P. Centricity accepted as a determinant property until a game similar to LoL, called Smite~\cite{hirez2014-smite}, was found that is ego-referenced rather than world-referenced. LoL only supports pausing under strict conditions (\eg in tournament play) and so could be said to support \nZ. Persistence is the only criterion left as the ``defining characteristic with which to distinguish between video games and virtual worlds''~\cite{girvan2013-virtualworld}.

In the analysis, Diablo and LoL were classified as pseudo-persistent worlds.
Pseudo-persistence is a powerful tool that can be used to blur the lines between video games and virtual worlds \eg through the use of instances. In the game called Destiny~\cite{bungie2014-destiny,truman2015-destiny}, the universe is created by replacing the virtual lobby, of Diablo and LoL, with `seamless background matchmaking'~\cite{truman2015-destiny}. The universe is made up of instances called `bubbles', which connect with `Activity Hosts' and a `WorldServer'. A single bubble, is standardized to simultaneously support six players and twenty-five software agents; Activity Hosts are cloud-based machines that run missions logic; and the WorldServer persists all character and progression data, using cloud computing~\cite{truman2015-destiny}. A bubble is hosted seamlessly on either a player's console (a `private' bubble) for peer-to-peer play or hosted in the cloud (a `public' bubble), if it is a public event. Destiny's architecture is still only pseudo-persistent and not a virtual world according to Section~\ref{section:shared_spatiality_within_one_shard}. The WorldServer can not be judged as a virtual world, since the spatiotemporal {\VE}s are the bubbles. And, the bubbles are not virtual worlds either, since they are copies of space and lack persistence.

\subsubsection{Transhumance}

\citeN{soderlund2009-proximity} describes how a pseudo-persistent world can be created using ad-hoc networking \eg using the Transhumance platform, where the collection (a `fog', rather than a `cloud') of devices forms the world. The Transhumance middleware platform concentrates heavily on security and power consumption~\cite{paroux2007-transhumance}, but is rich enough to implement the pervasive game called Team Exploration~\cite{demeure2008-transhumance}.

\citeN{paroux2007-transhumance} do mention a `time-limited mode' for group creation, but do not go into detail on how time is implemented \eg how time is synchronized between all devices. It is assumed that \ST is used, pending research pertaining to \VT on MANets. Each user has their own device, so no turn-taking is needed \ie satisfying \Rt. The spatial representation of Team Exploration is an allegory of physical space, because it models that space. A player's location in the virtual is mapped to their physical location and could be depicted on a virtual map. Even though the player must move through physical space to move virtually, it can not be denied that \VS is present. Team Exploration supports \OS, because all players of the game share the same virtual space. Transhumance provides for \Sh, where the fog of devices can form a common shared data space~\cite{paroux2007-transhumance}. If nodes are out of reach, network partitions may occur where no route can be found between two nodes~\cite{demeure2008-transhumance}. 
As long as the common data space does not become partitioned, the Transhumance platform forms \ONE shard, else each partition becomes a separate shard.
Each device has its own data space which can be considered a temporary instance when not merged. Further changes, on previously merged devices, that have become partitioned are not part of the shard unless synchronized again. Game mechanics can be put in place from a cultural perspective, to entice players to meet, so that data spaces merge~\cite{nevelsteen2015-pervasivemoo} \eg in Transhumance, all members of the team had to be network connected to each other (in proximity) to solve a puzzle~\cite{demeure2008-transhumance}. 
No mention of \SA was found in the references, but it is assumed that it would not be impossible to implement them. In the absence of \VT, agents would have to be activated according to \ST (\ie mixed reality). Because \SA is not supported, \Ix is not satisfied either. Transhumance is assumed to use \ST, satisfying \nZ.
During a pervasive gaming session, the platform can provide for a pseudo-persistent world through the fog of devices. Individually, each device in the fog is vulnerable to failure (not supporting \dP), but Transhumance enables a degree of persistence by having users of a group host data for other members of the group~\cite{paroux2007-transhumance}. And, lastly, nothing like an avatar was found in reference to Team Exploration; a player's own location is not shown on the map and neither the location of other players connected to their device.

Because \VT is mapped to physical spatiality, and \ST is assumed to be used, that would classify the Transhumance platform as a mixed reality technology (see Section~\ref{section:mixed_reality} below) \ie with mixed reality interaction (not \Ix) . If \VT was present, \SA implemented with \Ix, and players provided with \Av, the Transhumance platform could provide for a pseudo-persistent mixed world. 

\todoin{
Determining the criteria needed to make a MANet support a virtual world is also a contribution?!
}

\subsubsection{Virtual Reality} 
\label{section:virtual_reality}

\todoin{
The concept of virtual reality has perhaps shift over the last two decades, from including immersive virtual reality and cyberspace~\cite{mclellan1994-vr}, to what 

is \cite{mclellan1994-vr} already cited?
}

\citeN[p.v]{coquillart2011-vr} describe virtual reality as ``interactive human-computer-mediated simulations of artificial environments''. Only some of those artificial environments will be virtual worlds according to Definition~\ref{def:vw}, leading to the statement, by \citeN{dionisio2013-metaverse}, that virtual worlds ``constitute a subset of virtual reality applications''.

\subsubsection{Mixed Reality} 
\label{section:mixed_reality}

\citeN{milgram1999-mixed} present a continuum for mixed reality, between the virtual and physical. \citeN{benford1998-boundaries} describe mixed reality as a shared space that integrates the dimensions of the local and remote, and the physical and synthetic. If \VT, \VS, \SA, \Ix or \Av have been `mapped' to a physical counterpart, then the world should fall on the continuum in between the pure virtual and physical. Note, that because WoW makes use of both \ST (\eg for the day/night schedule) and \VT, to be precise, that would make WoW not purely virtual \ie a mixed world.

\subsubsection{The Metaverse}
\label{section:metaverse_beyond}

According to Definition~\ref{def:vw}, the Internet fails to be a virtual world, failing the criteria of a \OT, \Rt, \Sh and \P. Because the Internet has multiple spatiotemporal shards, the Internet is perhaps more closely related to a `world of worlds'~\cite{bartle2010-history} than one world. \citeN{dionisio2013-metaverse} call a future topology for multiple virtual worlds `metagalaxies' or the `Metaverse'. The key difference between the Internet and the Metaverse, is that the Metaverse would support \Rt. Because the Internet is already mixed reality (\eg with video conferencing, web cameras depicting a live video feed of cities in the physical world, tele-operations, and projections from the net onto buildings), it is possible to conclude that the Metaverse will be necessarily mixed as well.

\subsubsection{The Matrix}

By entertaining the concept found in the movie The Matrix~\cite{wachowski1999-matrix}, Definition~\ref{def:vw} can be tested to see if it is future proof.
If Neo takes the `red pill' and finds the real world is actually a simulation, is the real world a virtual world according to Definition~\ref{def:vw}? To start, the current real world can be renamed to the Matrix. The `new' world that Neo discovers can then be called the `real' world. Matrix time is then an abstraction of time equal to that of \VT. Two human agents, Neo and Trinity, accessing the Matrix from the their real world, share real-world time and Matrix time (\VT), satisfying \OT. The Matrix supports \Rt and is \Sh\footnote{A MUD, inside the Matrix, is then contained in a shard nested inside the shard of the Matrix. The users in the Matrix are segregated from the users inside MUD and users are prohibited from `physically' moving between the shard and nested shard.
}%
(as we know of today). 
The Matrix has \SA; Neo and Trinity can interact with each other in the Matrix, and act and react with the Matrix itself; Matrix time is \nZ; the Matrix supports \P; and each agent of the Matrix is given a \Av.
In addition, the Matrix offers ego-referenced centricity, and a tethered referenced view (or perhaps a world referenced view for those who have an `outer body experience'). The Matrix is rendered directly in the minds of its users, Neo and Trinity.
The communication architecture the Matrix runs on, is left as the fiction depicted in the movie.

%% file: incl/7-conclusion.tex
\sectioncaps{Conclusion}

The primary result of this article is a detailed definition for a virtual world, with all underlying terms defined. The definition is applied directly to technology for classification, showing which properties set apart the different technologies. Remaining properties which do not determine a virtual world, are listed so they can be used to distinguish between virtual worlds. The implication of this article is to provide a definition that is detailed enough to be adopted by the research community. The novelty in using grounded theory to classify technologies that implement a virtual world, is that if a new technology emerges it can be handled. If the new technology challenges the current theory, properties can be added to the theory and the definition updated.

Herein a definition has be derived for a technology that \ISA virtual world. 
As future work, perhaps a definition can be derived for a technology that \HASA virtual world \eg by relaxing the \nZ criterion (technology perhaps \mbox{\HASA} virtual world when a player has control over time through pausing) or using the properties of centricity or presence. It could be argued that a game such as Civilization~V~\cite{firaxis2010-civ5} \HASA virtual world wherein a player moves games pieces, mimicing a board game. Perhaps \citeN{bartle2010-history} specified the uniqueness criterion for a `virtual self' in an attempt to delineate when a technology is \ISA virtual world, classifying technologies with \MANY virtual selves as the likeness of a board game that \HASA virtual world. In addition, a qualitative analysis characterizing the worldliness of a virtual world could prove useful \eg how many is `\MANY' or do more believable \SA contribute to the worldliness of world?

%% file: incl/Appendix.tex
\appendix

\section*{Appendix}
\addcontentsline{toc}{section}{Appendix}
\setcounter{section}{1}
\label{section:appendix}

\begin{inparaenum}[(I)]
\item MUD~\cite{trubshaw1978-mud} and \item World of Warcraft (WoW)~\cite{blizzard2004-wow} were immediately added to the sampling set, a characteristic first generation virtual world and a characteristic contemporary massive multiplayer virtual world, respectively.
Thereafter, technologies were added that would separate the digital world from the virtual \ie a \item calculator~\cite{ti1993-ti36x} and \item graphics calculator~\cite{ti2004-ti84plus}. 
\item Email chess~\cite[Correspondence chess]{wikipedia2016-org} was added, because it is a two player game, involved turn taking and also included networking. 
MUD is text-based, so a single user word processing application was added \ie \item Microsoft Word~\cite{microsoft1983-word}; and, also a word processor utilizing networking for collaborative editing \ie \item Google Docs (GDocs)~\cite{google2007-docs}. To differentiate on static or animated drawing, \item a CAD application~\cite{autodesk2010-maya} was added to the sampling set, taking into consideration with and \item without animation. \item A telephone call~\cite[Telephone]{wikipedia2016-org} and \item television~\cite[Television]{wikipedia2016-org} were added, due to the discussion by \citeN{combs2004-def}. This led to the addition of a virtual variant of the telephone, \item \citeN{skype2003}, and also the question of how chat compares \ie \item IRC~\cite{kalt2000-irc}. To differentiate MUD from its predecessor, \item Zork~\cite{supnik1994-zork} was added.  
To differentiate between the various virtual worlds, \item Ultima Online~\cite{originsys1997-uo}, \item Big World Technology~\cite{bigworld2002} and \item EVE Online~\cite{ccpgames2003-eve} were added; the first two because they used shards and regions, respectively, and the latter because of its architecture and science fiction outer space theme.
\item Diablo~\cite{blizzard1996-diablo} was added due to it being a video game and it being discussed in \cite{combs2004-def}; the contemporary variant (which also uses a virtual lobby) \item Leagues of Legends (LoL)~\cite{riot2009-lol} was also added. 
\item Elite: Dangerous~\cite{frontier2014-elite} was added, because of its use of instances.
As a graphical modern variant to turn-taking email chess, \item Civilization V (Civ5)~\cite{firaxis2010-civ5} was added. \item Doom~\cite{id1993-doom} was added because it was a classic first-person shooter in 3D. 
The social network called \item \citeN{facebook2004} was added because it is discussed in \cite{bell2008-expanded}. The addition of \item Second Life~\cite{linden2003-secondlife}, is due to it being a virtual social community. \item Stunt.io~\cite{sundsted2012-stunt} was added because it is a virtual world engine displaying web pages similar to Facebook. And, \item the Internet~\cite[Internet]{wikipedia2016-org} as a whole must be considered. 
Some technologies were eventually dropped from the results table (\eg a telephone call and Diablo), because their resulting classification was identical to others in the study.
\end{inparaenum}

\paragraph{A Note on Scripting} 

Some of the technologies in the study have scripting languages available for extending them (\eg IRC, MUD). Given a sufficiently rich scripting language, it might prove impossible to determine the limits of their capabilities. Scripting is one technique through which a technology can be repurposed beyond its intended use~\cite{nevelsteen2015-pervasivemoo} \eg if a web server is enhanced with scripting to serve as a game server, is it still just a web server? In the study, technologies are considered without what is ultimately implementable if their scripting language was to be employed.

\subsection{Virtual (Simulated) Environment (\VE)}

All technologies listed in Table~\ref{table:results}, except for the calculator, implement some form of \VE. A calculator is a digital technology that simulates calculation on a digital display; the abstraction is not high enough to be considered implementing a \VE.

\subsection{Shared Temporality (\OT)}
\label{section:appendix_ot}

In email chess, players do not necessarily have to share any of the mentioned temporalities (real-world (\texttt{R}), hardware (\texttt{H}), simulated (\texttt{S} or \ST) or virtual (\texttt{V} or \VT)), but can maximally share real-world and \ST; players play on different computers, with no \VT implemented, only turn-taking. %
The digital hardware of a calculator lacks both \ST and \VT. Since such a calculator is operated by just a single person, that person shares real-world and hardware time, with themselves and the calculator.
CAD software, loaded with a CAD drawing, runs according to \ST \eg the CAD software can rotate the drawing according to certain clock speeds. If the CAD drawing contains animation, then subjective time and an epoch of zero to the start of the animation can be part of the loaded state \ie support for \VT.
GDocs only uses real-world and \ST; actions are translated into timed events that are synchronized on a cloud platform and disseminated to the various clients.

Skype is similar to a telephone call in the sense that participants share real-world time, but also records objective real-world times as \ST, synchronizing %
\begin{wrapfigure}{r}{0.5\textwidth}
\begin{adjustbox}{scale=0.8}

\ttfamily\selectfont
\begin{tabular}{l|c|c|cl|}

\textsc{technology}
						& \ST   	    
								& \VT
										& \OT	
												& 			\\
\hline                                  
email chess				& \Y	& \N	& \R	& \EQ RS	\\
calculator				& \N 	& \N 	& \R	& \EQ RH	\\
CAD drawing				& \Y 	& \N 	& \R	& \EQ RHS	\\
CAD/Maya animation		& \Y 	& \Y 	& \Y	& \EQ RHSV	\\
Google Docs				& \Y	& \N	& \R	& \EQ RS	\\
\hline                                  
Skype incl.~video	 	& \Y	& \N	& \R 	& \EQ RS	\\
IRC						& \Y	& \N	& \R	& \EQ R[S]	\\
\hline                                  
Zork					& \Y	& \N	& \R	& \EQ RH[S]	\\
Civ5, localhost			& \Y	& \Y	& \N	& \EQ RHS 	\\
Civ5, networked			& \Y	& \Y	& \Y	& \EQ RSV	\\
Civ5, single player		& \Y	& \Y	& \Y	& \EQ RHSV	\\
Civ5, email				& \Y	& \Y	& \N	& \EQ RS	\\
Doom (Deathmatch)		& \Y	& \Y	& \Y	& \EQ RHSV(RSV)\\
League of Legends		& \Y	& \Y	& \Y	& \EQ RSV	\\
\hline                                  
MUD, localhost			& \Y	& \Y	& \Y	& \EQ RHSV	\\
MUD						& \Y	& \Y	& \Y	& \EQ RSV	\\
WoW, \ONE shard 		& \Y	& \Y	& \Y	& \EQ RSV	\\
WoW, all shards			& \Y	& \Y	& \N	& \EQ RS[V+]\\
\hline                                  
Facebook				& \Y	& \N	& \R	& \EQ RS	\\
the Internet			& \Y	& \Y	& \N	& \EQ R[SV+]\\
\hline

\end{tabular}
%\caption{partial table}
\label{table:sharedtime}

\end{adjustbox}
\end{wrapfigure}
these times to the different Skype clients; \VT is not supported. Although the IRC protocol does, have a call to query another system's \ST (marked \texttt{[S]}), the IRC chat protocol~\cite{kalt2000-irc} is an order list of events that is irrelevant of time, as long as they are kept sequentially intact. IRC does not implement \VT, only real-world time is shared.

Although Zork has potential access to \ST (\texttt{[S]}), it is a player's actions which progress the game state. This lacking is apparent in Zork not supporting \SA (see Section~\ref{section:software_agents}). %
Civ5 is a turn-based game that can be played against a human or Artificial Intelligence (AI) agent. When played on a localhost (in `Hotseat' mode), players share real-world, hardware and \ST. They do not, however, share \VT; each player plays a time slot of \VT and then passes control to the other player. If played through a network (over LAN or via a dedicated server), players no longer share hardware time, but they do share \ST and \VT in principle \ie the player is playing, but just not allowed to make any actions (the \Rt property will account for the lack of actions). In a single player game, when playing against an AI agent, the game behaves similarly; the player is playing, but not allowed to make any actions during the AI agents turn.
It is also possible to play Civ5 over email, meaning Civ5 inherits the same \OT aspects as email \ie maximally sharing real-world and \ST~\cite{cawley2010-civ5}, even though \VT is implemented.
Doom supports \VT because the game state can be frozen, saved and loaded. Doom and MUD share all temporalities in a single player localhost mode. Doom in multiplayer (max four) `Deathmatch' mode, LoL, MUD (when played through a network) and WoW, share all temporalities except for hardware temporality.
In WoW, there is a reference to real-world time in-game and some in-game effects are related to real-world time \eg like the day/night schedule. %
When considering all the shards of WoW (see Section~\ref{section:one_shard}), not one, but many {\VT}s are supported \ie it is assumed that only one \ST is synchronized across all WoW servers.

Facebook is equal to Skype, with shared real-world time and a \ST across cloud computing.
The Internet is not one system such as Facebook, but many, supporting many {\ST}s and {\VT}s (each system on the Internet potentially having its own synchronized clock, plus abstractions); not all agents share the same \VT, because there are \MANY (\texttt{+}).

\subsection{Real-time (\Rt)}

The real-time property only serves to classify turn-based technologies, such as email chess and Civ5. The Internet in its entirely includes email and so can not be considered a real-time system.

\subsection{Shared Spatiality (\OS)}

It can be argued that players control the pieces on the email chess board in a space supporting spatial properties \ie supporting an allegory of space (\VS). Players of email chess share \VS, supporing \OS.

Each digital digit on the display of a calculator is set in the display space, but the calculator does not simulate an allegory of physical space. %
CAD drawings are often models of the physical world, and so do support \VS.
A CAD drawing is a single user environment, so \VS is shared with the user and the other entities in the space. 
GDocs might do a great job implementing an allegory of a sheet of paper, but not of physical space. For a calculator and GDocs, a representation of the user is not contained within the space.

Skype offers chat and also supports video: the chat doesn't provide \VS on the same grounds that GDocs doesn't. Actual physical space is portrayed in video, but participants are viewing other participants moving through physical space (which is not modeled, see Section~\ref{section:ve}), rather than virtual space.

Zork and MUD are peculiar, because they \textbf{do} implement \VS, but without a visualization of space; the allegory of space comes from the text description of each room~\cite[p.81]{qvortrup2002-virtualspace}, player movement and topology. 
A MUD with one room would be the equivalent to a chat service. When compared to MUD, it %
\begin{wrapfigure}{r}{0.5\textwidth}
\begin{adjustbox}{scale=0.8}

\ttfamily\selectfont
\begin{tabular}{l|c|c|c|}

\textsc{technology}		
									& \VS
											& \OS
													& \Sh	\\
\hline                                                      
email chess							& \Y	& \Y	& \Y 	\\
calculator							& \N	& \R	& \Y	\\
CAD drawing							& \Y	& \Y	& \Y	\\
Google Docs							& \N	& \R	& \Y	\\
\hline                                                      
Skype incl.~video	 				& \N	& \R	& \Y	\\
IRC									& \N	& \R	& \N	\\
\hline                                                      
Zork								& \Y 	& \Y	& \Y	\\
Civ5								& \Y 	& \Y	& \Y	\\
Doom (Deathmatch) 					& \Y 	& \Y	& \Y	\\
League of Legends					& \Y 	& \Y	& \Y	\\
\hline
sim/MUD, \ZERO players	 			& \Y 	& \Y	& \Y	\\
MUD, localhost						& \Y 	& \Y	& \Y	\\
MUD, \ONE room						& \N 	& \R	& \Y	\\
MUD									& \Y 	& \Y	& \Y	\\
WoW, \ONE shard						& \Y 	& \Y	& \Y	\\
WoW, all shards						& \Y	& \R	& \N	\\
Big World (regions, dynamic)		& \Y 	& \Y	& \Y	\\
EVE Online (regions, static)		& \Y 	& \Y	& \Y	\\
\hline                                                      
Facebook							& \N	& \R	& \Y	\\
Stunt.io							& \N	& \R	& \Y	\\
the Internet						& \Y	& \R	& \N	\\
\hline

\end{tabular}
%\caption{partial table}
\label{table:sharedspace}

\end{adjustbox}
\end{wrapfigure}
can be argued that IRC also allows a user to move between `channels' using text commands. However, IRC still lacks an allegory of physical space, which MUD provides through narrative. 
The \OS of Zork is based on a single user environment, like a CAD drawing.
\VS is supported by Civ5, Doom, LoL and WoW; they have an allegory of physical space with the likeness of a Euclidean 3D space. Civ5, Doom, LoL, MUD (with multiple rooms) and a single shard of WoW, support \OS. 

Facebook does not implement \VS on the same grounds that GDocs does not. To complicate matters, Stunt.io~\cite{sundsted2012-stunt} is a virtual world engine that acts as a web server; it does fail to provide \VS if it just displays static web pages. 
If the entire Internet is considered, \VS will be implemented somewhere.

\subsection{\ONE Shard (\Sh)}

In email chess, the game state consists of one or more data spaces spread out on one or more computers holding the email; the game state, however, is \Sh. When players email their moves to the other player(s), the game state is synchronized so that the next move can be made. %
The calculator, CAD drawing and Zork only have one data space and so are \Sh.
GDocs and Facebook~\cite{pingdom2010-facebook} are powered by cloud computing; many computers, each with their own data space that are synchronized to form the state of one document, news feed, profile, \textit{etc}. %

Skype, Civ5, Doom and MUD, can be considered similar to email chess \ie partial state held in potentially more than one data space that is synchronized to form \Sh. 
IRC is chat service that spans multiple servers, with groups of servers supporting \MANY IRC networks~\cite{irchelp2014-networks}; each IRC network can be considered a different shard. 

Although all are video games, LoL is very different than Civ5 and Doom, in how their data spaces are set up. LoL has a networked virtual `lobby' where players gather, that want to play a game against each other. When a group of players has been formed, they enter a data space separated from other players. When the game is over, they are returned to the lobby. According to Section~\ref{section:shared_spatiality_within_one_shard}, each game is an instance similar to those of WoW. When results of each game are synchronized with the player profiles and leaderboards in the global data space, one \Sh is formed.

WoW is pronounced to be divided into a number of shards (called `realms'), each with various zones and land masses (mini worlds). There are many data spaces in WoW, but each `realm' is \Sh served by a single server; the entirety of WoW, all shards considered, is multiple shards. 
WoW has a feature where a group of players can play an \textit{instance} dungeon; a data space where only the designated group of players are playing together, similar to a single game of LoL. When the instance dungeon is over, results of the instance dungeon are synchronized with the shard. In recent years, upgrades to WoW have allowed players from different shards to play in the same instance dungeon together; results from the instance dungeon are synchronized with each respective shard, but shards do not synchronize. 
BigWorld is \Sh, but contrary to WoW, space is dynamically regionalized. EVE Online is \Sh that is statically regionalized; one (SOL) server per solar system~\cite{url:emilsson2010-infinitespace}.

Similar to the entirety of WoW, all shards considered, the Internet is multiple shards. To access the Internet, a user uses a browser to connect to any number of different shards, by typing in an address.

\subsection{Size}

Text-based MUDs had considerably less movement space than some modern worlds, but were virtual worlds nonetheless; movement speeds have been set so that it takes tens of minutes to run through Doom or fly through WoW.
In Doom, the player is teleported from one `level'~\cite[p.~693]{gregory2014-v2} to the next and teleporters also exist within a single level. Within one shard of WoW, there are mini worlds (called islands or planets, according to narrative) and a player can travel continuously across one mini world, but must teleport between mini worlds. As with Doom, teleporters exist in WoW within mini world.

\subsection{Indoor/Outdoor}

The setting for Doom is various levels set in factory-like complexes, with occasional areas where the player is let `outdoors'. However,  outdoors is a box that restrains the player's movement, but in a way that is believable with respect to the game. A level is set within a `sky box'~\cite{gregory2014-v2} which is superimposed with an image of an outdoor skyline. A sky box is usually the outermost box in the environment; beyond the sky box aspects of the environment are usually undefined. 
MUD has a room defined for each point a player a can move to. Indoor and outdoor is a matter of narrative. If the player were to be able to escape the room structure of MUD, the area would be undefined.  
The graphical backdrop of WoW makes it mostly an outdoor experience, with some buildings in cities and dungeons being depicted as indoor experiences. 
In WoW, areas have either an edge that limits player movement or an area that is detrimental to the player (\eg avatar death), stimulating them to backtrack.

\subsection{\MANY Human Agents (\HA)}

\begin{wrapfigure}{r}{0.4\textwidth}
\begin{adjustbox}{scale=0.8}

\ttfamily\selectfont
\begin{tabular}{l|c|c|c|}

\textsc{technology}		& \OH 
								& \TwoH 
										& \HA \\
\hline
email chess				& \Y	& \Y 	& \Y \\
% calculator			& \Y 	& \N 	& \Y \\
% CAD drawing			& \Y 	& \N 	& \Y \\
Zork					& \Y 	& \N 	& \Y \\
sim/MUD, \ZERO players	& \N 	& \N 	& \N \\
MUD, localhost			& \Y 	& \N 	& \Y \\
\hline

\end{tabular}
%\caption{partial table, resulting in H$^\PLUS$}
\label{table:oneplus}

\end{adjustbox}
\end{wrapfigure}

All technologies support \ONE human agent (\OH), except for a pure simulation of a world. Email chess is usually played by two or more (\TwoH), but you can email yourself.
A calculator, CAD drawing and Zork were designed to support a \OH.
GDocs supports \HA, with a maximum of 50 simultaneous users~\cite{url:google2015-docs}. Skype and IRC conversations, are often between two people, but can be a conference with multiple. 
Civ5 and Doom were designed with a single and multiplayer mode, supporting \HA; Doom has a campaign in which only single player is supported and also a networked `Deathmatch' competitive mode for up to four players. 
LoL supports \HA in a single game, but supports a more massive amount of players being connected to their servers simultaneously in the `lobby'. 
MUD supports \HA \eg active MUDs often have hundreds of simultaneous players on one server. 
If MUD is used without networking, then it allows for \ONE local player to be active~\cite{bartle2010-history}.
WoW supports massive amounts of players, with players spread over multiple shards.
Facebook deals with massive amounts of users and is a website on the Internet \ie the Internet also supports \HA.

\subsection{\MANY Software Agents (\SA)}

Email chess, a calculator, a CAD drawing (without animation), GDocs, Skype, IRC, Zork, and Facebook all do not support \VT \ie fail to support \SA.
A CAD drawing with animation supports \VT, but does not implement \SA.
IRC supports bots that automate channel maintenance, but these are not agents; they only react to incoming messages and do not exhibit continuous behavior.
In version 3.2B of Zork~\cite{supnik1994-zork} there is a troll located in the cellar; it is not an agent, but equal to that of an IRC bot.
Civ5, Doom, LoL, MUD and WoW all support \SA. 
Since Stunt.io is a virtual world engine, it does support \SA.
The Internet as a whole can not be discounted of \SA \ie somewhere on the Internet software agents will exist. 

\subsection{Virtual Interaction (\Ix)}

\begin{wrapfigure}{r}{0.4\textwidth}
\begin{adjustbox}{scale=0.8}

\ttfamily\selectfont
\begin{tabular}{l|c|c|c|}

\textsc{technology}				
						& \Sh				
								& \SA
										& \Ix	\\
\hline                                  
CAD/Maya animation		& \Y	& \N 	& \R	\\
sim/MUD, \ZERO players	& \Y	& \Y 	& \Y	\\
WoW, \ONE shard			& \Y	& \Y 	& \Y	\\
WoW, all shards			& \N	& \Y 	& \R	\\
Facebook				& \Y	& \R	& \R	\\
Stunt.io				& \Y	& \Y	& \Y	\\

\hline

\end{tabular}
%\caption{partial table}
\label{table:agentinteraction}

\end{adjustbox}
\end{wrapfigure}

Email chess, a calculator, a CAD drawing with and without animation, GDocs, Skype, IRC, Zork and Facebook do not support \SA \ie fail to support \Ix.
Some of these (\eg \mbox{Facebook}) allow \Ix between \HA (\eg communication tools), but without \SA, human agents can't have \Ix with the world, only act on it.
All the remaining technologies named in this subsection allow for \Ix with people and action/reaction with things and the environment.
In Civ5, players are confronted with artificial opponents and must have \Ix with their units. The monsters in Doom take prescribed against actions against the player allowing for \Ix. Players of LoL have \Ix with the other players, but the majority of units, which can be interacted with, are controlled by the game. MUD and WoW have entities with scripted behaviors. If all of WoW and the Internet is considered, the prerequisite of \Sh is not met. 
Facebook lacks the prerequisite of \SA. 
Stunt.io serves web pages and does support \SA. If no web interface is built such that the \SA can take action on the user without it being a reaction, then Stunt.io falls in the same category as a pure simulation of a world, where \SA only have \Ix with each other. 

\subsection{non-Pausable (\nZ)}

Email chess, Civ5 and the Internet fail to meet the \Rt prerequisite.
A calculator, GDocs, Skype, IRC, Zork and Facebook are all \nZ. 
A CAD drawing with animation can be paused, failing \nZ.
Doom supports \VT, but can be paused in both single player and Deathmatch multiplayer mode.
LoL supports pausing of the game, but only in tournament play (marked as positive for the benefit of the doubt). %
MUD and WoW both support \VT and are \nZ; if all the shards of WoW are taken into account, they collectively fail \OT, but still support \nZ.

\subsection{Persistence (\P, world (\wP), data (\dP)) and Pseudo-}

Email chess is a persistent technology, with both \dP and ($\land$) \wP; \dP is provided for by the email server and loading the world of email is just as persistent as loading the world of WoW.
A calculator has no support for \dP and so also fails \wP.
GDocs, Skype and IRC are persistent technologies.
IRC defeats the \dP criterion by not having much data to persist; there is no real-world data beyond the preservation of some nicknames \ie conversations are ephemeral. 

Doom does support \dP; it is possible to persist data, albeit manually. 
If Doom were left running to support \wP, it would fail \dP, if no one was around to save the state manually~\cite{nevelsteen2016-persistence}. So, it must fail either \wP exclusively or (\XOR) \dP.
The persistence of a CAD drawing, Zork, Civ5 and Diablo in single player, is equal to that of Doom.

For multiplayer Diablo, player(s) are first gathered in a virtual lobby, before venturing into dungeons. Each dungeon is similar to an instance dungeon of WoW, where the dungeon is created for one run through it and reset. Afterwards, player data is preserved and the player(s) can choose to start another dungeon. Instances fail \wP, but the virtual lobby is a persistent technology (the virtual lobby does fail other properties \eg \VS or \SA).
Because world data can be injected into an instance and results gathered from it, Diablo multiplayer is an example of pseudo-persistence. Pseudo-persistence can account for statements like: ``I think the persistance [\sic] of the avatar is more important than the persistance [\sic] of the world. Hence Diablo is a VW, while Starcraft isn't''~\cite[\#18]{combs2004-def}. The persistence of LoL, is equal to that of Diablo.

\citeauthor{bartle2003} co-created MUD and so it is assumed that his criteria was based on his own experiences with MUD \ie MUD passes his own criteria. 
MUD was designed to be accessed through networking, but it is possible to access a MUD from a local machine as a single player game~\cite{bartle2010-history}; under this condition, MUD is similar to Doom in the sense that if the game or the system is turned of, the world fails to exist~\cite{nevelsteen2016-persistence}. The difference being that, if MUD is left on for \wP, the world data is still persisted to storage periodically, supporting \dP.

A type of MUD, that has been referred to as a Groundhog Day MUD~\cite{keegan1997-classification}, is one where the entire world is reset periodically. Such a world would be equal to that of Diablo (\ie lacking \wP), unless something in the world has been under the constant influence of the world's time, in which case \wP and \dP would be required for that something~\cite{nevelsteen2016-persistence}.

WoW supports both \wP and \dP, except for instance dungeons (similar to the games of LoL), but which are synchronized to the shard.

Facebook supports \wP and \dP for user data. Community oriented data can be considered world data and is also persisted \ie Facebook is a persistent technology. 
The Internet, as a whole, satisfies \citeauthor{bartle2003}'s criterion for \wP. Property \dP for all parts of the Internet is problematic (\eg failing servers and services); \dP is marked as positive for the benefit of the doubt.

\subsection{\MANY Avatar (\Av)}

A calculator and a CAD drawing do not support \Av.
GDocs uses a 2D image to represent the presence of a user in a shared document, with a indicator in the document to show where they are and which changes to the document they are making \ie GDocs can not be discounted for supporting \Av. 
Skype makes use of a 2D image on a profile and in chat to represent the user. Interaction via chat is similar to that of MUD, but with text and other media shown to be emanating from a users image icon. Skype can be said to support \Av, on the same grounds that MUD does.
IRC is similar to MUD (and Skype), and so is marked as supporting text-based \Av.
Although similar to MUD, determining if Zork makes use of an avatar is more problematic. Zork is single player, with an ego-referenced viewpoint in the sense that the player can not see themselves and a 3rd-person perspective can not be used to see if the player has an avatar. But, the player has an entity in the game, with an inventory, that interacts with the fictional world and that is referred to as `you' in the game \eg ``The thief attacks, and you fall back desperately''. On the basis of interaction, the conclusion is that Zork supports \Av.
If pieces on a chess board are considered under the control of the player, players of both email chess and Civ5 control multiple entities. The question is whether those entities are avatars; it can be argued that those entities are just pieces on a game board with abilities and the player has no avatar. On the grounds that a player should identify with their virtual selves (a personification if you will), those entities are discounted from being avatars. In Civ5, each player must also select to play a famous world leader, which does satisfy \Av. Email chess is marked as not supporting \Av and Civ5 as supporting.
In a single player game of Doom, the player has an ego-referenced viewpoint, with an entity \SA can interact with \ie support for \Av. In Deathmatch mode, Doom makes use of 2D images (simulating 3D models) as entity representations for player opponents.
MUD uses text-based \Av, while LoL and WoW both use a 3D model as \Av.
According to \citeN{bell2008-expanded}, a Facebook profile explicitly does not qualify as an avatar, because of lack of agency. But, since all interaction with other agents is done through the user profile, Facebook is considered to support \Av here.
On the Internet as a whole, a user has many different representations \ie not one unique avatar, but \MANY; avatars are usually isolated to a particular shard.